\documentclass[paper]{agujournal2019}
\usepackage{url}
\usepackage{lineno}
\usepackage[inline]{trackchanges} 
\usepackage{soul}
\usepackage{color}
\usepackage{amsmath}
\usepackage{amssymb}
\usepackage{graphicx}

\draftfalse

\newcommand{\planss} {{Planetary Space Science }}  
\newcommand{\ssr}{   {Space Sci. Rev. }}

\newcommand{\jgr}{   {J. Geophys. Res.}}
\newcommand{\grl}{   {Geophys. Res. Lett.}}

\newcommand{\apjl}{   {Astrophys. J. Lett.}}

\graphicspath{{./}{figures/}}

\journalname{Geophysical Research Letters}

\begin{document}

\title{{Picturing global substorm dynamics in the magnetotail using low-altitude ELFIN measurements and data mining-based magnetic field reconstructions}}

\authors{Xiaofei Shi \affil{1}, Grant K. Stephens \affil{2},  Anton V. Artemyev \affil{1}, Mikhail~I. Sitnov \affil{2}, Vassilis Angelopoulos \affil{1}}
\affiliation{1}{University of California, Los Angeles, Los Angeles, CA, USA}
\affiliation{2}{Applied Physics Laboratory, Johns Hopkins University, Laurel, MD, USA}

\correspondingauthor{Xiaofei Shi}{sxf1698@g.ucla.edu}

\begin{keypoints}
\item {Electron and ion isotropy boundaries (IBe,i) inferred from ELFIN particle data are compared with IBs derived from empirical magnetic field}
\item {ELFIN and empirical magnetic field IBs consistently move equatorward in the growth phase and diverge in latitude after the substorm onset}
\item {ELFIN and empirical magnetic field IBe reveal similar V-like patterns associated with local accumulation of magnetic flux in the tail}
\end{keypoints}

\begin{abstract}
{A global reconfiguration of the magnetotail characterizes substorms. Current sheet thinning, intensification, and magnetic field stretching are defining features of the substorm growth phase and their spatial distributions control the timing and location of substorm onset. Presently, sparse in-situ observations cannot resolve these distributions. A promising approach is to use new substorm magnetic field reconstruction methods based on data mining, termed SST19. Here we compare the SST19 reconstructions to low-altitude ELFIN measurements of energetic particle precipitations to probe the radial profile of the equatorial magnetic field curvature during a 19~August 2022 substorm. ELFIN and SST19 yield a consistent dynamical picture of the magnetotail during the growth phase and capture expected features such as the formation of a thin current sheet and its earthward motion. Furthermore, they resolve a {\it V}-like pattern of isotropic electron precipitation boundaries in the time-energy plane, consistent with earlier observations but now over a broad energy range.}
\end{abstract}

\section*{Plain Language Summary}
The solar wind strongly stretches the magnetic field on the night side of our planet. This stretching slowly accumulates and rapidly relaxes during special periods called substorms. These variations are difficult to investigate because of the extreme sparsity of spacecraft observations. The problem can be solved by properly sorting historical data from similar substorm phases to form swarms of synthetic probes which are then used to reconstruct the magnetic field configuration. {The low-altitude ELFIN mission provides another means of probing the magnetic field configuration by measuring electrons and ions, which become scattered as the night side magnetic field is stretched. Here, we demonstrate that both approaches yield a consistent picture of the night side magnetic field and how it changes during a substorm that occurred on 19~August 2022.}

\section{Introduction}\label{sec:intro}
Substorms are one of the most energetic phenomena in Earth’s magnetosphere, responsible for a large-scale reconfiguration of the magnetotail, charged particle acceleration and injection into the inner magnetosphere, as well as the formation of strong field-aligned currents that couple the magnetosphere to the ionosphere~\cite{Baker96,Angelopoulos08,Sitnov19}. During the substorm growth phase, the magnetotail magnetic field stretches significantly and a thin current sheet forms ~\cite<e.g.,>[and references therein]{Sergeev12,Runov21:jastp}. Magnetic reconnection in this thin current sheet enables acceleration of charged particles and their injection into the inner magnetosphere \cite<see reviews by>[and references therein]{book:Gonzalez&Parker,Sitnov19,Birn21:AGU}. 
In-situ spacecraft measurements provide important information for investigating substorm dynamics. However, given its substantial volume, even multi-spacecraft missions cannot resolve the large-scale magnetotail structure and its evolution during substorms~\cite<see discussion in>{Stephens19,Sitnov19:jgr}.
{Nevertheless, two general approaches exist that can resolve the global-scale structure of the magnetosphere from data. The first approach leverages machine learning algorithms to reconstruct magnetospheric plasma quantities from sparse in-situ datasets~\cite<e.g.,>{Sitnov08,Bortnik16}. The second involves remote sensing the magnetosphere using ground-based observations~\cite<e.g.,>{MenkWaters13} or low-altitude orbiting spacecraft~\cite{Dubyagin02,Coxon18}. In recent years, both approaches have been employed to progress the understanding of large-scale substorm dynamics. For instance, by applying a data mining (DM) technique to more than two decades of space magnetometer data, \citeA{Stephens19} reconstructed the dynamics of the magnetotail magnetic field during substorms. Termed SST19, their algorithm revealed}
the formation of multi-scale current sheets, the stretching and dipolarization of the tail magnetic field, the 3D structure of the substorm current wedge, and the location of magnetic reconnection verified by in-situ observations of the ion diffusion region ~\cite{Stephens23}. 

{Remote sensing can be exemplified by the Electron Losses and Fields Investigation (ELFIN), a pair of twin CubeSats launched in September 2018 into a polar low-Earth orbit to observe the precipitation of charged particles with magnetospheric origins~\cite{Angelopoulos20:elfin}}.
{Low-altitude spacecraft, such as ELFIN, allow the} global magnetotail spatial profile to be inferred~\cite<e.g.,>{Sergeev11,Sergeev18:grl,Sergeev23:elfin}. This approach relies on the isotropization of proton and electron distributions when the magnetic field curvature becomes comparable to or smaller than the particle gyroradius resulting in chaotization of their orbits and pitch angle scattering~\cite{Sergeev&Tsyganenko82}. It reveals more of the large-scale magnetotail structure and its reconfiguration during substorms than magnetic field times-series measurements from (approximately stationary) equatorial satellites which lack spatial information of the magnetic field profile ~\cite<e.g.,>{Sergeev12:IB,Sergeev15, Dubyagin02,Artemyev22:jgr:ELFIN&THEMIS}. The basic elements of such a reconfiguration include the formation of a thin current sheet with magnetic field line stretching and earthward current sheet motion~\cite{Wanliss00,Kozelova&Kozelov13,Petrukovich07,Artemyev16:jgr:thinning}. Current sheet thinning scatters energetic ions and electrons into the loss cone causing them to precipitate into the ionosphere. Low-altitude measurements of precipitating particles of various species and energies can localize the latitude of this scattering (from which the location of the thin current sheet, its Earthward-most edge, and the magnetic field radius of curvature which depend on the particles' energy and species can be inferred) thus tracing the thin current sheet's dynamics~\cite{Yahnin97,Sergeev12:IB,Sergeev18:grl,Sivadas17}. 

So far, comparisons between empirical magnetic field reconstructions and low-altitude measurements have been made using event-oriented models~\cite<e.g.,>{Kubyshkina09,Kubyshkina11,Sergeev23:elfin}. In that approach, a statistical model of the magnetic field, fit to a large archive of historical magnetometer data~\cite<e.g.,>{Tsyganenko95}, was additionally tweaked to achieve better consistency with low-altitude precipitation patterns using a small number of {lucky (and hence rare)} observations made during the event {and in the region of interest, that is close to IBs}. In contrast, the SST19 DM approach employed in this study reconstructs the magnetic field using an ``event-oriented" subset of the archive ($\approx~1\%$), which is still large enough ($\sim 9\cdot 10^4$ synthetic probes) to allow for a far more flexible magnetic field architecture and an increased sensitivity to storm and substorm variations. Here we {compare the DM-based magnetic field reconstruction of the magnetotail during a 19~August 2022 substorm to ELFIN   measurements of energetic ions and electrons~\cite{Angelopoulos20:elfin}. This is a unique substorm event in that ELFIN passed through latitudes that map to the near-Earth ($r\approx5$--$20 R_E$) magnetotail about midnight magnetic local time (MLT) six times.}
{This comparison demonstrates that SST19 and ELFIN reveal a consistent global dynamical picture of the magnetotail during substorms.}

\section{Spacecraft Observational dataset}\label{sec:data}
We use the {low-orbit-altitude} ($\sim 450$ km) twin ELFIN CubeSats (A and B) to measure energetic electron ($50$keV--$6$MeV) and ion ($250$keV--$5$MeV) fluxes~\cite{Angelopoulos20:elfin}. ELFIN moves along a polar orbit with a period of $\sim1.5$ hour and measures particle fluxes with an angular resolution of $\sim 22.5^\circ$ and a temporal resolution of 2.8~s (spin period) \cite{Angelopoulos20:elfin}. In this study, we use two types of ELFIN data products: energy spectra of locally precipitating (within the loss-cone) and locally trapped (outside the loss-cone) fluxes; only measurements with more than five counts per bin (energy, time) are included \cite<see details of data products in>{Angelopoulos23:ssr}. 

We use the $\textit{SML}$ and $\textit{SMR}$ indices from the SuperMag project~\cite{Gjerloev09} to monitor substorm and storm activity. {The Supporting Information (SI) also contains a comparison of the low-altitude ELFIN measurements with near-equatorial energetic particle fluxes from the THEMIS~\cite{Angelopoulos08:sst} and Magnetospheric Multiscale (MMS)~\cite{Burch16} missions. The THEMIS Solid State Telescope (SST)~\cite{Angelopoulos08:sst} provides electron fluxes of $30-700$~keV at a $3$~s time resolution. The MMS Fly's Eye Energetic Particle Spectrometer (FEEPS) \cite{Blake16} provides electron and ion fluxes of $50-650$keV at a $2.5$~s time resolution.}

\begin{figure*}
\centering
\includegraphics[width=0.8\textwidth]{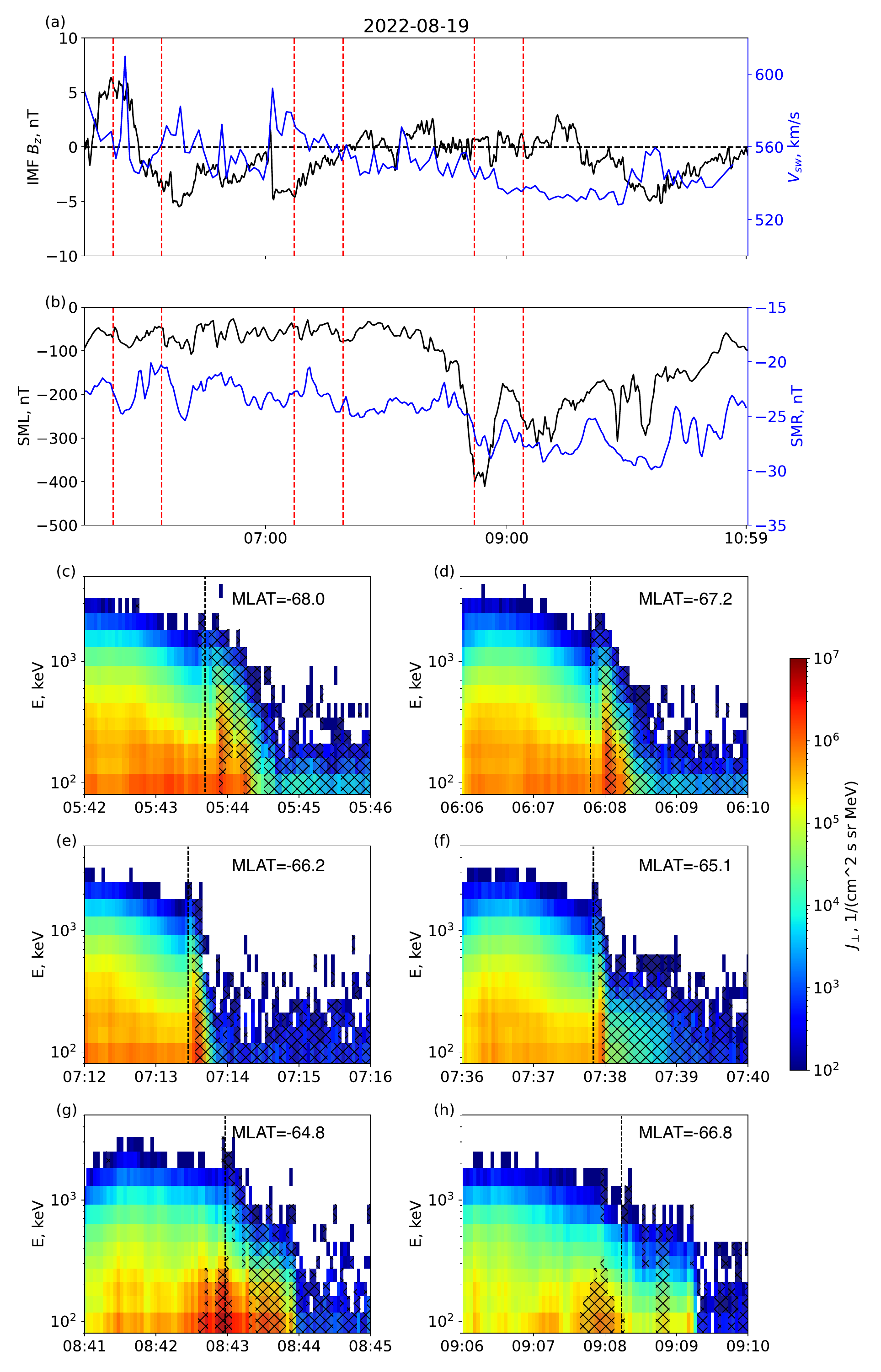}
\caption{Overview of the substorm activity from 05:00--11:00 UT on 19~August 2022. {(a) Interplanetary magnetic field and solar wind velocity measured by ARTEMIS-P1 (THEMIS probe B) which was located within the solar wind;} (b) SuperMag substorm, $\textit{SML}$ (black), and storm, $\textit{SMR}$ (blue), indices; (c--h) ELFIN's observations of electron energy spectra for locally trapped fluxes for six orbits intersect the midnight sector (the crossing times are indicated by the red dashed lines in panels (a) and (b)), and the isotropic regions are shaded on top. {Panels (c, e, g) are observations from ELFIN A while panels (d, f, h) are from ELFIN B.}
\label{fig1}}
\end{figure*}

{Figure~\ref{fig1}(a) shows the solar wind magnetic field ($B_z$ in the Geocentric Solar Magnetospheric system or GSM) and the solar wind velocity from 05:30 to 11:00~UT on 19~August 2022 \cite<measured by THEMIS B; see>{Angelopoulos11:ARTEMIS}: there is a prolonged interval ($\sim$06:00--08:00~UT) with $B_z<0$, that corresponds to the magnetosphere loading during substorm growth phase. Figure~\ref{fig1}(b) shows the $\textit{SML}$ and $\textit{SMR}$ indices.} The consistently moderate negative value of $\textit{SMR}$, $\sim -20$~nT, suggests the presence of a ring current ion population in the inner magnetosphere. This negative value remains relatively stable, indicating continuous replenishment of this population by new injections \cite<e.g.,>{Gkioulidou14}. The {dip} of $\textit{SML}$ around 08:40~UT indicates a substorm onset, and thus before this moment, there should be magnetotail current sheet thinning ({the growth phase begins with the southward turning of the IMF around 06:00~UT}). Six ELFIN passes intersect the near-midnight sector during this interval; red vertical lines in Figure \ref{fig1}(a,~b) mark those times. Figures \ref{fig1}(c--h) show ELFIN's observations of electron energy spectra for locally trapped fluxes ($J_{\perp}$), with the shading indicating where the fluxes are nearly isotropic (where the trapped flux is comparable to the precipitating flux, $J_{\perp}\sim J_{\parallel}$). The energy spectra exhibit typical characteristics of various magnetospheric regions \cite<see>{Mourenas21:jgr:ELFIN,Angelopoulos23:ssr}. In Figure \ref{fig1}(c), for example: (1) from 05:42:00 to 05:43:50 UT ELFIN crossed the outer radiation belt which is characterized by fluxes of relativistic electrons ($>500$~keV) with strong flux anisotropy ($J_{\perp}\gg J_{\parallel}$). (2) From 05:43:50 to 05:44:35 UT, a decrease in energy levels is observed for electrons with $J_{\perp}>10^3/cm^2/s/sr/MeV$, while the fluxes are mostly isotropic ($J_{\perp}\sim J_{\parallel}$). This is the so-called isotropy boundary (IB) \cite{Imhof77, Sergeev&Tsyganenko82, Wilkins23}, the transition region between the outer radiation belt and the plasma sheet. The energy-time spectra there exhibit the energy/latitude (time) dispersion characteristic of the IB, with the latitude of the minimum energy of isotropization increasing with decreasing energy (the beginning of this transition region is indicated by the vertical line with the magnetic latitude, $\textit{MLAT}$, shown).  (3) From 05:44:35 to 05:46:00~UT, fluxes are isotropic and limited to within $<200$ keV energy; this is the plasma sheet region \cite{Artemyev22:jgr:ELFIN&THEMIS}. {The electron isotropy boundary (IBe) separates the plasma sheet and the outer radiation belt} and can be considered as an inner (equatorward) edge of the magnetotail plasma sheet. Note that the location for ion IB (IBi) and electron IB (IBe) are quite different, with IBi equatorward of the IBe due to the much larger gyroradius (and field-line curvature radius responsible for isotropization) of ions than electrons of the same energy \cite{Sergeev12:IB}. 

The location and shape of IBs vary during a substorm \cite<e.g.,>{Sergeev12:IB,Wilkins23}. Figure \ref{fig1}(c--h) reveals the dynamics of the IBe. During the growth phase (Figure \ref{fig1}(c--f)), the magnetic latitude $|\textit{MLAT}|$ of the IBe's equatorward edge decreases, indicating that the IBe moves equatorward. After substorm onset, (Figure \ref{fig1}(h)), the $|\textit{MLAT}|$ of the IBe's equatorward edge increases, indicating that the IBe moves further poleward. The ELFIN orbit shown in Figure \ref{fig1}(g) crossed the magnetotail shortly after substorm onset, but the $|\textit{MLAT}|$ of the IBe kept decreasing. {To explain why we examine THEMIS observations near the plasma sheet (Figure~S1): from 06:00 to 07:10 UT, there is a decrease in the equatorial $B_z$ and an increase in $|B_x|$, indicating magnetotail current sheet thinning during the substorm growth phase \cite{Sergeev11,Artemyev16:jgr:thinning}, and between 07:10 and 08:50~UT, the THEMIS satellites recorded a sequence of dipolarizations, characterized by $B_z$ increases (and perturbations) and $|B_x|$ decreases (Figure~S1a). However, in between such dipolarization, THEMIS observed periods where $|B_x|$ increased and $B_z$ decreased, indicating there was continuous current sheet thinning but interrupted by (or interleaved with) several dipolarizations. Such short time-scale and likely localized dynamics of the magnetotail current sheet are not resolved in the $\textit{SML}$ profile (a global index) that shows a single substorm onset around 08:40UT. 
Thus, this is a complex substorm with enhanced dissipation occurring in the plasma sheet before the major onset (see discussions of similar phenomena in~\cite{Shukhtina2014,Yahnin2001}), but includes multiple ELFIN passes near midnight, which is infrequently observed, making this event particularly interesting. Such complex substorm dynamics does not allow to identify if the ELFIN orbit from Figure \ref{fig1}(g) crossed the stretched or dipolarized magnetotail configuration.}

{We also estimate ELFIN's position by comparing its particle fluxes during intervals of isotropic flux measurements, with the flux data from MMS and THEMIS spacecraft (Figure~S1), the method tested recently in \cite{Artemyev22:jgr:ELFIN&THEMIS,Shen23:jgr:ELFIN_dropout}. This comparison demonstrates that the IBe observed by ELFIN maps to near $10R_E$, the likely transition region between the plasma sheet and the outer radiation belt \cite<in agreemnet with>{Sergeev12:IB} at that time (more details are provided in the SI).}

\section{Data mining-based magnetic field reconstruction}\label{sec:model}

Here we compare the ELFIN observations, particularly the inferred IB locations, with those derived from an empirical magnetic field constructed using the DM-based algorithm SST19~\cite{Stephens19}. SST19 differs from conventional empirical geomagnetic field models~\cite<e.g.,>[and refs. therein]{Tsyganenko&Sitnov05} in two critical aspects. First, it describes the magnetospheric magnetic field using several sets of basis functions rather than custom-made modules. The number of such basis functions, used for the description of the magnetic field generated by equatorial and field-aligned currents as well as their shielding currents on the magnetopause, can be increased to resolve important morphological features such as the eastward ring current~\cite{Stephens16} or the Harang discontinuity~\cite{Sitnov2017M}. Second, SST19 employs a DM algorithm~\cite{Sitnov08} to fit such a flexible and multi-parameter magnetic field architecture to data. This exploits the recurrent nature of storms and substorms to augment the handful of space-borne magnetometer observations available at the moment of interest with a much larger set of observations (the nearest neighbors or NNs) made when the magnetosphere was in a similar storm/substorm configuration, based on the geomagnetic indices, their time-derivatives, and the strength of the solar wind driving. The number of NNs, $k_{NN}$, must be large enough, $k_{NN}\gg1$, to avoid over-fitting, while at the same time small enough, $k_{NN}\ll k_{DB}$ where $k_{DB}\sim10^{7}$ is the whole database of historical magnetometer records since 1995, to make the reconstructions sufficiently sensitive to the specific phases of substorms and storms. The distinct features of SST19, compared to its storm-time predecessor, TS07D~\cite{Tsyganenko&Sitnov07,Sitnov08}, are the use of substorm indices $\textit{AL}$ or (most recently) $\textit{SML}$~\cite{Gjerloev2012} and their time derivatives, as well as two independent basis function descriptions for thick and thin (presumably ion-scale) current sheets. The buildup and decay of the latter is a key feature of the magnetospheric reconfiguration during substorms~\cite{Sergeev11}. {More specifically, here we utilize the ``merged resolution" version of SST19, which concurrently resolves both the inner magnetosphere and the magnetotail \cite{Stephens&Sitnov21}. SST19 has been extensively validated by comparing its reconstructed magnetic field to the field observed by in-situ spacecraft~\cite{Stephens19,Sitnov19:jgr,Stephens&Sitnov21,Sitnov21,Stephens23}. The only notable modifications to SST19 employed here are an updated formulation of the thin current sheet spatial structure and an increased quantity of MMS magnetometer data.} More details of SST19 are provided in the SI. 

\begin{figure*}
\centering
\includegraphics[width=22pc]{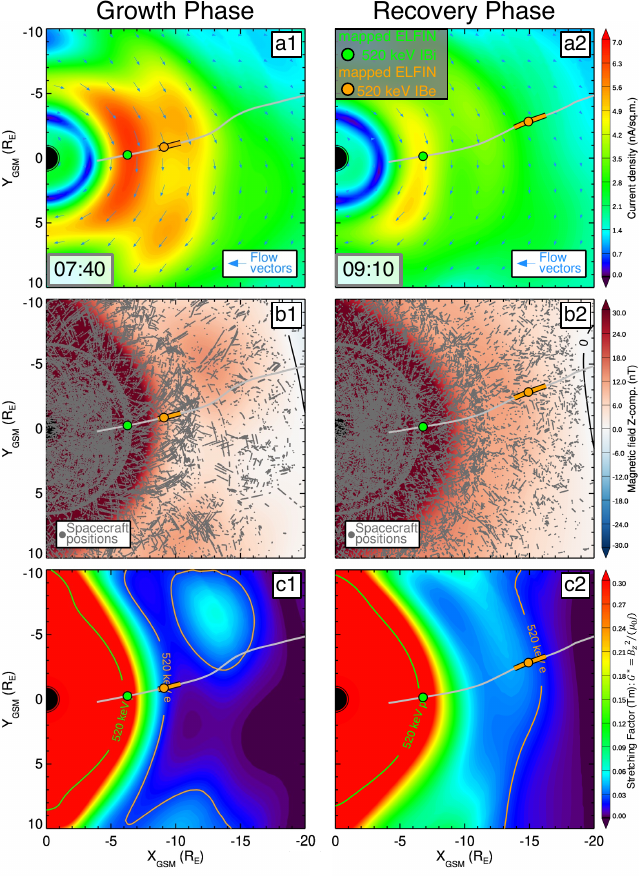}
\caption{{SST19 data mining-based reconstructions of the 19~August 2022 substorm, the left column is during the substorm growth phase (07:40~UT) and the right is during the recovery phase (09:10~UT): (a) Color-coded equatorial distributions of the electric current density, $j$, with arrows overplotted to indicate the direction of vector current density, $\mathbf{j}$. (b) Color-coded equatorial distributions of the z-component of the magnetic field, $B_z$, with grey dots overplotted to indicate the locations, projected to the $x$-$y$ plane, of the spacecraft magnetometer observations identified using the $K_\mathrm{NN}$ procedure and used to fit the analytical description of the magnetic field. (c) Color-coded equatorial distributions of the stretching factor $G^*=B_z^2/(\mu_0 j)$. The critical values of this parameter corresponding to the IB for 520~keV electrons and protons are shown by the orange and green contours respectively.
The location of ELFIN mapped to the magnetic equator when it observed the 520~keV IBi and IBe are overplotted in green and orange circles respectively. The corresponding error bars quantify the uncertainty of the IB determination using $J_{||}/J_{\perp}$ distributions as described below in Figure~\ref{fig3}. Grey lines show projections of the ELFIN orbits to the equatorial plane. 
\label{fig2}}}
\end{figure*}

The ELFIN-observed IBs for species $\gamma=e,p$ and the associated particle rigidities $G_{\gamma}=m_{\gamma}V_{\gamma}/e$ (where $m_{\gamma}$ are the masses of electrons and protons, and $V_{\gamma}$ their velocities), can be compared to those derived from empirical magnetic field reconstructions by mapping their positions to the magnetic equator and computing {there} the equivalent parameter $G^*=B_z^2/(\mu_0j)$, termed the stretching factor ($B_z$ is the northward component of the equatorial magnetic field and $j$ is the current density), {as is demonstrated in Figure~\ref{fig2}}. According to~\citeA{Sergeev&Tsyganenko82} and \citeA{Sergeev18:grl}, the transition to isotropy due to chaotization of particle orbits occurs when $G^*<8G$. {Figures~\ref{fig2}a--\ref{fig2}c} show the equatorial distributions of {$j$, $B_z$, and} $G^*$ {respectively} for {two} moments {during the substorm growth (07:40~UT) and recovery phase (09:10~UT) corresponding to the times of the fourth and sixth ELFIN passes indicated by the vertical red lines} shown in Figure~\ref{fig1} {(all six passes at 05:45, 06:10, 07:15, 07:40, and 08:45~UT are shown in Figures~S2 and S3)}. 
Over-plotted on these panels are the
mappings of the ELFIN's position {when it observed IBe and IBi at the energy 520~keV (the middle energy channel in the range of 0.1-1 MeV). Note that these mappings nicely match the corresponding IB contours derived from SST19.} 

Figures~S4 and S5 show the equatorial $B_z$ and $G^*$ mapped to ELFIN's altitude on an $\textit{MLT}$-$\textit{MLAT}$ grid. These plots facilitate the subsequent comparison of ELFIN-observed IBs with those inferred from the SST19 magnetic field reconstruction. {Also, to allow the comparison of these reconstructions with similar results from other missions~\cite<e.g.>{Sergeev18:grl}, the analogs of Figures~S4 and S5 in coordinates $MLT$ and $AACGM$~\cite{Shepherd2014} are provided in Figures~S6 and S7. In addition, Figure~S8 presents validation of the SST19 reconstructions using the observed magnetic field from THEMIS and MMS.}

{Figure~\ref{fig2} reveals several important substorm features resolved by SST19. During the growth phase, a strong ($>5$~nA/m$^2$) current forms in the near-tail from $r\approx5$--$13 R_E$ (Figure~\ref{fig2}a1) accompanied by a $B_z$ minimum ($B_z=4.5$~nT) around $11 R_E$ (Figure~\ref{fig2}b1). This stretches the near-tail, as indicated by the non-monotonically decreasing $G^*$ distribution, which possesses a local minimum about $11 R_E$ in the pre-midnight sector (Figure~\ref{fig2}c1). The tail configuration dramatically changes during the expansion phase, which persists into the recovery phase, signified by the collapse of the cross-tail current (Figure~\ref{fig2}a2) and a dipolarization of the magnetic field (Figure~\ref{fig2}b2). This inflates the value of $G^*$ across most of the tail (Figure~\ref{fig2}c2), pushing the reconstructed IBe and IBi to larger radial distances. Note that the domain over which the SST19 reconstructions are presented here is limited to $r \leq 20 R_E$. Beyond this limit the mapping may be strongly complicated by magnetic reconnection~\cite{Stephens23}.}



\section{Comparison of ELFIN observations with DM reconstructions}\label{sec:compare}

\begin{figure*}
\centering
\includegraphics[width=1.1\textwidth]{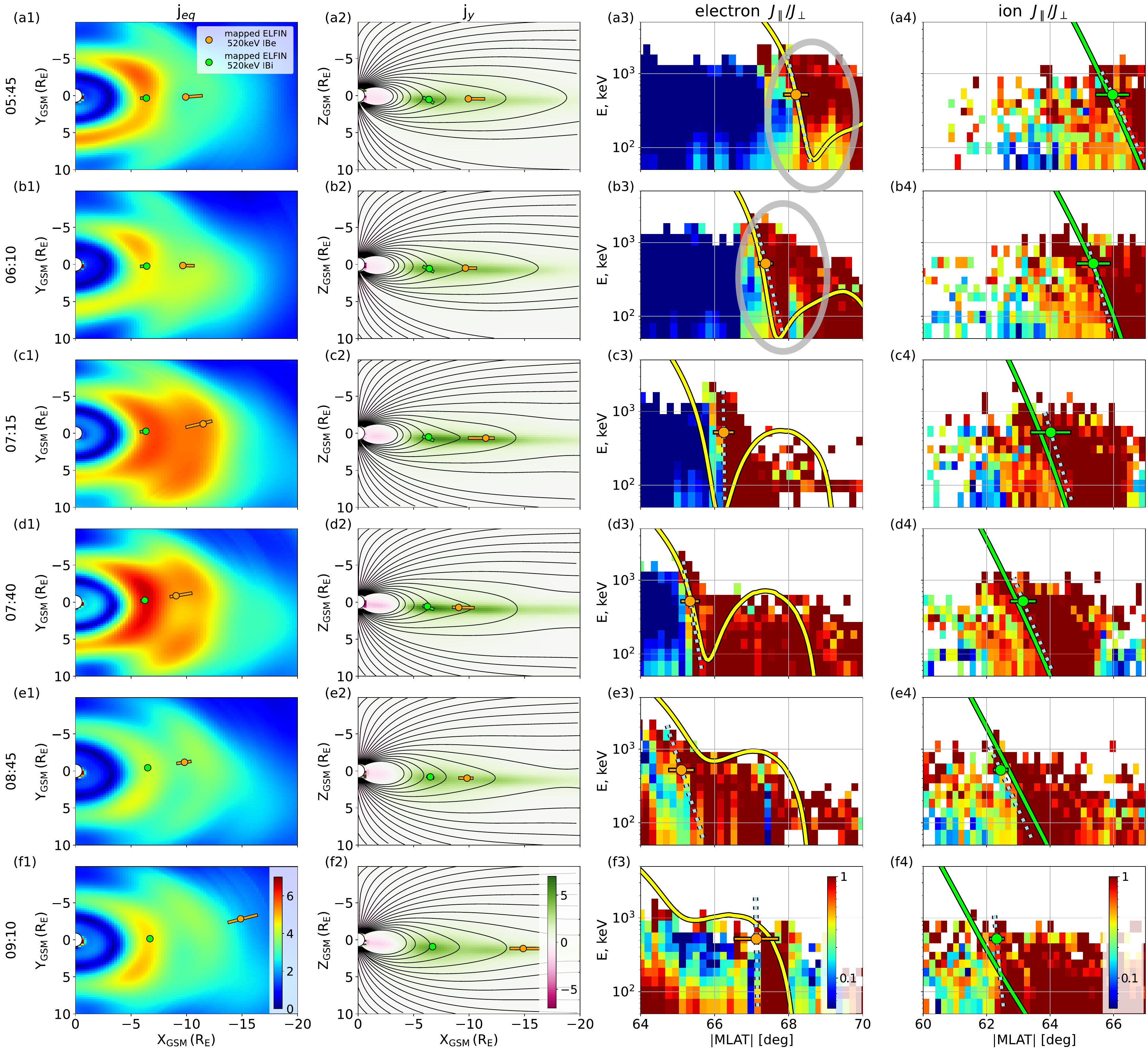}
\caption{An overview of the SST19 magnetic field reconstructions and ELFIN observations for {six} sub-intervals. The SST19 results show equatorial (Column 1) and Y-GSM components (Column 2) of the cross-tail current density, respectively, projected on the GSM planes shown in the axes. The equatorial projections of the $520$~keV electron and ion isotropy boundaries (IB) determined by ELFIN (traced to the equator using SST19 magnetic field reconstruction) are shown by colored dots {(the horizontal bars show the uncertainty of IB observations)}. Various times are depicted in various rows {(a-f)}. $J_{\parallel}/J_{\perp}$ for electrons (Column 3) and ions (Column 4) measured by ELFIN around the times corresponding to rows (a-f) in the left two columns. {Blue dotted curves represent the IB locations as determined by ELFIN observations. The orange and green dots indicate the positions of the $520$~keV IB, with horizontal bars illustrating the associated uncertainties. Yellow and green curves depict the electron and ion isotropy boundaries (IBe and IBi) from the SST19 reconstructions, respectively.} {Gray ellipses in Panels (a3, b3)} highlight {\it V}-like IBe patterns arising from SST19 reconstruction that bears similarity to the pattern of precipitating-to-trapped flux ratio observed by ELFIN at that time.
\label{fig3}}
\end{figure*}

{Figure~\ref{fig3} provides an overview of the SST19 magnetic field reconstructions and ELFIN observations for all six ELFIN orbits:  before the substorm growth phase, 05:40-05:45UT (panels a1-a4); during the substorm growth phase, and 06:06-06:10 UT, 07:10-07:15UT and 07:37-07:40 UT (panels b1-b4, c1-c4 and d1-d4); during the expansion phase of the substorm, 08:40-08:45UT (panels e1-e4); and during the recovery phase, 09:10-09:15 UT (panels f1-f4)}. The two left columns display the SST19 results, illustrating the equatorial and meridional distributions of the cross-tail current density. On the right-hand side, the two columns show the $J_{\parallel}/J_{\perp}$ ratios for both electrons and ions measured by ELFIN during the same time sub-intervals. The pre-substorm time is characterized by a weak cross-tail current density in the center plasma sheet ($<4$nA/m$^2$, panel a1) and a thick current sheet (panel a2). {The IBe is located around $|\textit{MLAT}|\sim 68^\circ$ with the $\Delta |\textit{MLAT}|\sim 0.5^\circ$ range for $[100,1000]$~keV, whereas IBi is around $|\textit{MLAT}|\sim 66^\circ$. During the substorm growth phase, the cross-tail current density increases to $>5$~nA/m$^2$ (panel c1) and is concentrated within the thin current sheet (panel c2). The IBe is moved earthward to $|MLAT|\sim 66.5^\circ$ (panel c3), and shrinks to $\Delta |\textit{MLAT}|\sim 0.1^\circ$ range for $[100,1000]$~keV.} Thus, for sub-intervals shown in panels (a, c) ELFIN observations of IBe and IBi are consistent with the SST19 reconstructions: stronger current density and thinner current sheet are associated with a smaller distance between IBe and IBi, and earthward motion of both boundaries {\cite<see>[for discussion of similar earthward IBs motions derived from POES measurements during the growth phase]{Sergeev12:IB}.} 

Figures~\ref{fig3}e and \ref{fig3}f describe a very short expansion phase and the following recovery phase. SST19 exhibits cross-tail current density distributions consistent with the expansion and recovery phases, characterized by weak current density and a broader current sheet (as seen in panels e1, f1 and e2, f2 compared to panels c1 and c2). {ELFIN shows the IBe at $|\textit{MLAT}|\sim 65^\circ$ and $|\textit{MLAT}|\sim 67^\circ$ (panels e3 and f3), and the latter returns to its value in the early growth phase at 06:10 UT, while IBi at $|\textit{MLAT}|\sim 62^\circ$ (panels e4 and f4), is found well equatorward of its growth phase location.} 

The third and forth columns of Figure~\ref{fig3} also compare the ELFIN precipitation ratios $J_{\parallel}/J_{\perp}$ with the IBs derived from SST19 and marked by {yellow and green lines for IBe and IBi, respectively}. The latter closely follow the ELFIN IBs seen as sharp transitions to dark red pixels (isotropization) {marked by blue dotted lines. To determine the IB position, we select all pairs of bins (two latitudinal bins) for fixed energy ($E$) where the flux ratio crosses from $\geq 0.6$ to $< 1.0$. These groups of points in the energy and latitudinal space are then fitted by a power-law function $|\textit{MLAT}| = a\cdot E^b$ \cite<the power-law $MLAT-E$ fitting underlines a nonlinear relation between particle energies and equatorial magnetic field (radial distance) in the equation of pitch-angle scattering rate, see>{Birmingham84,Delcourt94:scattering}. This function indicates the IB position, while the standard deviation of the fitting describes the uncertainty range of the IB position. The resulting discrepancies between IBs derived from the merged resolution SST19 model and ELFIN data (except IBes in the expansion phase) are much smaller than $1^{\circ}$ MLAT, typical errors reported for statistical and adaptive models~\cite{Shevchenko2010}. Note that the latter, which somewhat outperform the former, cannot be applied to our event because of the absence of any real probes in the IB source region.} 

Interestingly, in Figures~\ref{fig3}a3 and \ref{fig3}b3, the SST19 IBes closely follow similar ELFIN IBes forming the characteristic V-like patterns {(first discussed in~\cite{Artemyev2023})}. They suggest the appearance of the dipolarized (less stretched) magnetic field region tailward of the left (lower-latitude, high-energy) IBe, where the electron precipitation is suppressed by the enhanced equatorial magnetic field (increasing the curvature radius). Similar non-monotonic profiles of the tail stretching parameter ($G^*$) were reported by~\citeA{Sergeev18:grl} based on the POES data for selected energies (30 and 100 keV). With ELFIN measurements we resolve such non-monotonic profiles within a wider energy range and show their consistency with similar SST19 features. More such IBe transitions at higher latitudes (in the plasma sheet, far tailward from IBe) are observed by ELFIN but not resolved by SST19, suggesting that ELFIN {\it has a potential to detect} even more complex patterns of alternating stretched and dipolarized regions in the tail~{\cite{Artemyev2023}}, which may have implications for magnetotail stability and dynamics $B_z$~\cite{Erkaev07,Pritchett&Coroniti2010,Sitnov10,Birn18}.

\begin{figure*}
\centering
\includegraphics[width=0.7\textwidth]{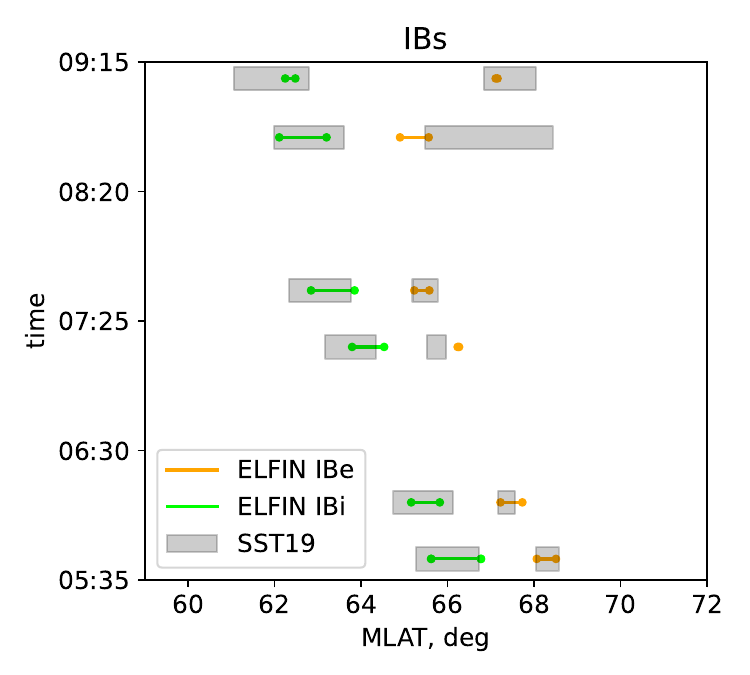}
\caption{Dynamics of isotropy boundaries for electrons and ions derived from the SST19 reconstructions (grey-shaded) and ELFIN measurements (orange for electrons and green for ions). {The energy range used to determine the $\textit{MLAT}$ range of isotropy boundaries is from 100~keV to 1~MeV.}
\label{fig4}}
\end{figure*}

To further quantify the comparison of ELFIN observations of IBs with SST19, we plot the latitudinal range of IBs for all six ELFIN orbits in Figure \ref{fig4}. 
During the entire interval, 05:40--09:30, the SST19 IBe traces well the dynamics of IBs from ELFIN observations: {IBe (orange) moves equatorward (its equatorial projection moves earthward) during the growth phase. The overall dynamics of the IBi (green) and its inference from SST19 magnetic reconstruction are similar to those of the IBe during the growth phase. After the substorm onset, IBe and IBi diverge in latitude. Notably, SST19 traces well the motion of IBs during the substorm growth phase and after substorm onset.}


{
To ensure the reliability of the reconstruction algorithm, we provide in Figure~S9 (in a format similar to Figure~\ref{fig2}) the reconstruction of three more ELFIN observed IB events, one of which is another {\it V}-like pattern event shown in Figure~8 of \citeA{Artemyev2023} (2022-08-11). Yet another event (2022-08-07) shown first in Figure~8 of \cite{Artemyev2023}, occurred in the period of high geomagnetic activity when the consistency of ELFIN IBs with those derived from SST19 is not expected because precipitation is strongly modified by plasma waves.
}

Thus, Figures~\ref{fig3} and \ref{fig4} show reasonable overall consistency between the SST19 DM reconstructions and the ELFIN data. 
The residual differences, especially after the substorm onset, can be explained by an additional electron scattering caused by whistler-mode waves, that form intense precipitation bursts equatorward from the IBe (see, e.g. Figures \ref{fig3}e3 and \ref{fig3}f3) around $\textit{MLAT}\in[-64,-65^\circ]$) and increase the uncertainty range for IBe position determination \cite<see>[for detailed investigation of night-side electron precipitation events associated with whistler-mode wave activity]{Tsai22,Artemyev24}. Another possible cause of IBe variations are transient meso-scale perturbations of the equatorial magnetic field \cite<e.g.,>{ Lin14:hybrid_code,Panov&Pritchett18,Sorathia20}. Further systematic comparative analyses of DM reconstructions and ELFIN observations could clarify the roles of different transient precipitation mechanisms resulting in the observed spatial/temporal variability of the IBe.

\section{Conclusions} \label{sec:conclusions}
{In this study, we investigate the substorm dynamics of electron and ion isotropy boundaries (IBe and IBi), which are the transition regions between the inner edge of the electron plasma sheet and the outer radiation belt, and between the inner edge of the ion plasma sheet (the tail current sheet) and the ring-current, respectively. By combining low-altitude ELFIN measurements of energetic particle (ions and electrons) spectra with the SST19 DM-based empirical magnetic field reconstruction algorithm~\cite{Stephens19,Stephens23}, we demonstrate that multiple localized reductions of energetic electron precipitation within the IBe (seen as a {\it V}-like pattern) are associated with quasi-steady features in the magnetic field configuration reproduced by SST19. This comparison confirms the previously suggested interpretation of {\it V}-like patterns as the formation of a flux accumulation region in the near-Earth tail resulting in reduced field line stretching tailward of the transition region~\cite{Sergeev18:grl}. This flux accumulation region may be the locus of a developing magnetotail instability prior to substorm onset~\cite{Sitnov19}.
The overall agreement between the location of the ELFIN observed and SST19 reconstructed IBe and IBi are on the order of $\textit{MLAT} < 0.5^\circ$ over a broad range of energies ($100$--$1,000$~keV) during the substorm growth phase and includes their equatorward (earthward) motion. \citeA{Sergeev12:IB} have demonstrated similar current sheet dynamics by comparing POES IB measurements with a dynamically-adapted magnetospheric model \cite{Kubyshkina11} assimilating a rare conjunction of multiple THEMIS and GOES satellites. Following \citeA{Sergeev12:IB}, our results demonstrate the potential of combining low-altitude energetic particle measurements with magnetic field reconstruction algorithms for probing magnetotail substorm dynamics.}

\acknowledgments
We are grateful to NASA's CubeSat Launch Initiative for ELFIN's successful launch in the desired orbits. We acknowledge the early support of the ELFIN project by the AFOSR, under its University Nanosat Program, UNP-8 project, contract FA9453-12-D-0285, and by the California Space Grant program. We acknowledge the critical contributions of numerous volunteer ELFIN team student members and support by NASA 80NSSC22K1005 and NSF grants AGS-1242918, AGS-2019950. We acknowledge NASA contract NAS5-02099 for the use of data from the THEMIS Mission. The work of G. K. S. and M. I. S. was supported by NASA grants 80NSSC19K0074, 80NSSC20K1271, and 80NSSC24K0556. A.V.A and X.S. acknowledge support from the NASA grants 80NSSC20K1788, 80NSSC23K0108, 80NSSC24K0558. V.A. acknowledges support by NSF grants AGS-1242918, AGS-2019950 and NASA contract NAS5-02099.


\section*{Open Research} \noindent 
Fluxes measured by ELFIN are available in the ELFIN data archive https://data.elfin.ucla.edu/ in CDF format.\\ 
THEMIS dataset and summary plot are available in http://themis.ssl.berkeley.edu \\
MMS dataset and summary plot are available in https://lasp.colorado.edu/mms/sdc/public \\
SuperMag dataset are available in https://supermag.jhuapl.edu \\
The SST19 reconstruction output data is available on a Zenodo archive at \\ https://doi.org/10.5281/zenodo.11625067 \\
\noindent
Data analysis was done using SPEDAS V4.1 \cite{Angelopoulos19}. The software can be downloaded from {http://spedas.org/wiki/}.




\begin{thebibliography}{}

\bibitem [\protect \citeauthoryear {%
{Angelopoulos}%
}{%
{Angelopoulos}%
}{%
{\protect \APACyear {2011}}%
}]{%
Angelopoulos11:ARTEMIS}
\APACinsertmetastar {%
Angelopoulos11:ARTEMIS}%
\begin{APACrefauthors}%
{Angelopoulos}, V.%
\end{APACrefauthors}%
\unskip\
\newblock
\APACrefYearMonthDay{2011}{{\APACmonth{12}}}{}.
\newblock
{\BBOQ}\APACrefatitle {{The ARTEMIS Mission}} {{The ARTEMIS Mission}}.{\BBCQ}
\newblock
\APACjournalVolNumPages{\ssr}{165}{}{3-25}.
\newblock
\begin{APACrefDOI} \doi{10.1007/s11214-010-9687-2} \end{APACrefDOI}
\PrintBackRefs{\CurrentBib}

\bibitem [\protect \citeauthoryear {%
{Angelopoulos}%
\ \protect \BOthers {.}}{%
{Angelopoulos}%
\ \protect \BOthers {.}}{%
{\protect \APACyear {2019}}%
}]{%
Angelopoulos19}
\APACinsertmetastar {%
Angelopoulos19}%
\begin{APACrefauthors}%
{Angelopoulos}, V.%
, {Cruce}, P.%
, {Drozdov}, A.%
, {Grimes}, E\BPBI W.%
, {Hatzigeorgiu}, N.%
, {King}, D\BPBI A.%
\BDBL {}{Schroeder}, P.%
\end{APACrefauthors}%
\unskip\
\newblock
\APACrefYearMonthDay{2019}{{\APACmonth{01}}}{}.
\newblock
{\BBOQ}\APACrefatitle {{The Space Physics Environment Data Analysis System (SPEDAS)}} {{The Space Physics Environment Data Analysis System (SPEDAS)}}.{\BBCQ}
\newblock
\APACjournalVolNumPages{\ssr}{215}{}{9}.
\newblock
\begin{APACrefDOI} \doi{10.1007/s11214-018-0576-4} \end{APACrefDOI}
\PrintBackRefs{\CurrentBib}

\bibitem [\protect \citeauthoryear {%
{Angelopoulos}%
, {McFadden}%
\BCBL {}\ \protect \BOthers {.}}{%
{Angelopoulos}%
, {McFadden}%
\BCBL {}\ \protect \BOthers {.}}{%
{\protect \APACyear {2008}}%
}]{%
Angelopoulos08}
\APACinsertmetastar {%
Angelopoulos08}%
\begin{APACrefauthors}%
{Angelopoulos}, V.%
, {McFadden}, J\BPBI P.%
, {Larson}, D.%
, {Carlson}, C\BPBI W.%
, {Mende}, S\BPBI B.%
, {Frey}, H.%
\BDBL {}{Kepko}, L.%
\end{APACrefauthors}%
\unskip\
\newblock
\APACrefYearMonthDay{2008}{{\APACmonth{08}}}{}.
\newblock
{\BBOQ}\APACrefatitle {{Tail Reconnection Triggering Substorm Onset}} {{Tail Reconnection Triggering Substorm Onset}}.{\BBCQ}
\newblock
\APACjournalVolNumPages{Science}{321}{}{931-935}.
\newblock
\begin{APACrefDOI} \doi{10.1126/science.1160495} \end{APACrefDOI}
\PrintBackRefs{\CurrentBib}

\bibitem [\protect \citeauthoryear {%
{Angelopoulos}%
, {Sibeck}%
\BCBL {}\ \protect \BOthers {.}}{%
{Angelopoulos}%
, {Sibeck}%
\BCBL {}\ \protect \BOthers {.}}{%
{\protect \APACyear {2008}}%
}]{%
Angelopoulos08:sst}
\APACinsertmetastar {%
Angelopoulos08:sst}%
\begin{APACrefauthors}%
{Angelopoulos}, V.%
, {Sibeck}, D.%
, {Carlson}, C\BPBI W.%
, {McFadden}, J\BPBI P.%
, {Larson}, D.%
, {Lin}, R\BPBI P.%
\BDBL {}{Sigwarth}, J.%
\end{APACrefauthors}%
\unskip\
\newblock
\APACrefYearMonthDay{2008}{{\APACmonth{12}}}{}.
\newblock
{\BBOQ}\APACrefatitle {{First Results from the THEMIS Mission}} {{First Results from the THEMIS Mission}}.{\BBCQ}
\newblock
\APACjournalVolNumPages{\ssr}{141}{}{453-476}.
\newblock
\begin{APACrefDOI} \doi{10.1007/s11214-008-9378-4} \end{APACrefDOI}
\PrintBackRefs{\CurrentBib}

\bibitem [\protect \citeauthoryear {%
{Angelopoulos}%
\ \protect \BOthers {.}}{%
{Angelopoulos}%
\ \protect \BOthers {.}}{%
{\protect \APACyear {2020}}%
}]{%
Angelopoulos20:elfin}
\APACinsertmetastar {%
Angelopoulos20:elfin}%
\begin{APACrefauthors}%
{Angelopoulos}, V.%
, {Tsai}, E.%
, {Bingley}, L.%
, {Shaffer}, C.%
, {Turner}, D\BPBI L.%
, {Runov}, A.%
\BDBL {}{Zhang}, G\BPBI Y.%
\end{APACrefauthors}%
\unskip\
\newblock
\APACrefYearMonthDay{2020}{{\APACmonth{07}}}{}.
\newblock
{\BBOQ}\APACrefatitle {{The ELFIN Mission}} {{The ELFIN Mission}}.{\BBCQ}
\newblock
\APACjournalVolNumPages{\ssr}{216}{5}{103}.
\newblock
\begin{APACrefDOI} \doi{10.1007/s11214-020-00721-7} \end{APACrefDOI}
\PrintBackRefs{\CurrentBib}

\bibitem [\protect \citeauthoryear {%
{Angelopoulos}%
\ \protect \BOthers {.}}{%
{Angelopoulos}%
\ \protect \BOthers {.}}{%
{\protect \APACyear {2023}}%
}]{%
Angelopoulos23:ssr}
\APACinsertmetastar {%
Angelopoulos23:ssr}%
\begin{APACrefauthors}%
{Angelopoulos}, V.%
, {Zhang}, X\BPBI J.%
, {Artemyev}, A\BPBI V.%
, {Mourenas}, D.%
, {Tsai}, E.%
, {Wilkins}, C.%
\BDBL {}{Zarifian}, A.%
\end{APACrefauthors}%
\unskip\
\newblock
\APACrefYearMonthDay{2023}{{\APACmonth{08}}}{}.
\newblock
{\BBOQ}\APACrefatitle {{Energetic Electron Precipitation Driven by Electromagnetic Ion Cyclotron Waves from ELFIN's Low Altitude Perspective}} {{Energetic Electron Precipitation Driven by Electromagnetic Ion Cyclotron Waves from ELFIN's Low Altitude Perspective}}.{\BBCQ}
\newblock
\APACjournalVolNumPages{\ssr}{219}{5}{37}.
\newblock
\begin{APACrefDOI} \doi{10.1007/s11214-023-00984-w} \end{APACrefDOI}
\PrintBackRefs{\CurrentBib}

\bibitem [\protect \citeauthoryear {%
{Artemyev}%
, {Angelopoulos}%
, {Runov}%
\BCBL {}\ \BBA {} {Petrukovich}%
}{%
{Artemyev}%
\ \protect \BOthers {.}}{%
{\protect \APACyear {2016}}%
}]{%
Artemyev16:jgr:thinning}
\APACinsertmetastar {%
Artemyev16:jgr:thinning}%
\begin{APACrefauthors}%
{Artemyev}, A\BPBI V.%
, {Angelopoulos}, V.%
, {Runov}, A.%
\BCBL {}\ \BBA {} {Petrukovich}, A\BPBI A.%
\end{APACrefauthors}%
\unskip\
\newblock
\APACrefYearMonthDay{2016}{}{}.
\newblock
{\BBOQ}\APACrefatitle {{Properties of current sheet thinning at $x\sim 10$ to 12 $R_E$}} {{Properties of current sheet thinning at $x\sim 10$ to 12 $R_E$}}.{\BBCQ}
\newblock
\APACjournalVolNumPages{\jgr}{121}{}{6718--6731}.
\newblock
\begin{APACrefDOI} \doi{10.1002/2016JA022779} \end{APACrefDOI}
\PrintBackRefs{\CurrentBib}

\bibitem [\protect \citeauthoryear {%
Artemyev%
, Angelopoulos%
, Zhang%
, Chen%
\BCBL {}\ \BBA {} Runov%
}{%
Artemyev%
\ \protect \BOthers {.}}{%
{\protect \APACyear {2023}}%
}]{%
Artemyev2023}
\APACinsertmetastar {%
Artemyev2023}%
\begin{APACrefauthors}%
Artemyev, A\BPBI V.%
, Angelopoulos, V.%
, Zhang, X\BHBI J.%
, Chen, L.%
\BCBL {}\ \BBA {} Runov, A.%
\end{APACrefauthors}%
\unskip\
\newblock
\APACrefYearMonthDay{2023}{}{}.
\newblock
{\BBOQ}\APACrefatitle {Dispersed Relativistic Electron Precipitation Patterns Between the Ion and Electron Isotropy Boundaries} {Dispersed relativistic electron precipitation patterns between the ion and electron isotropy boundaries}.{\BBCQ}
\newblock
\APACjournalVolNumPages{Journal of Geophysical Research: Space Physics}{128}{12}{e2023JA032200}.
\newblock
\begin{APACrefURL} \url{https://agupubs.onlinelibrary.wiley.com/doi/abs/10.1029/2023JA032200} \end{APACrefURL}
\newblock
\begin{APACrefDOI} \doi{https://doi.org/10.1029/2023JA032200} \end{APACrefDOI}
\PrintBackRefs{\CurrentBib}

\bibitem [\protect \citeauthoryear {%
{Artemyev}%
\ \protect \BOthers {.}}{%
{Artemyev}%
\ \protect \BOthers {.}}{%
{\protect \APACyear {2022}}%
}]{%
Artemyev22:jgr:ELFIN&THEMIS}
\APACinsertmetastar {%
Artemyev22:jgr:ELFIN&THEMIS}%
\begin{APACrefauthors}%
{Artemyev}, A\BPBI V.%
, {Angelopoulos}, V.%
, {Zhang}, X\BPBI J.%
, {Runov}, A.%
, {Petrukovich}, A.%
, {Nakamura}, R.%
\BDBL {}{Wilkins}, C.%
\end{APACrefauthors}%
\unskip\
\newblock
\APACrefYearMonthDay{2022}{{\APACmonth{10}}}{}.
\newblock
{\BBOQ}\APACrefatitle {{Thinning of the Magnetotail Current Sheet Inferred From Low-Altitude Observations of Energetic Electrons}} {{Thinning of the Magnetotail Current Sheet Inferred From Low-Altitude Observations of Energetic Electrons}}.{\BBCQ}
\newblock
\APACjournalVolNumPages{Journal of Geophysical Research (Space Physics)}{127}{10}{e2022JA030705}.
\newblock
\begin{APACrefDOI} \doi{10.1029/2022JA030705} \end{APACrefDOI}
\PrintBackRefs{\CurrentBib}

\bibitem [\protect \citeauthoryear {%
{Artemyev}%
\ \protect \BOthers {.}}{%
{Artemyev}%
\ \protect \BOthers {.}}{%
{\protect \APACyear {2024}}%
}]{%
Artemyev24}
\APACinsertmetastar {%
Artemyev24}%
\begin{APACrefauthors}%
{Artemyev}, A\BPBI V.%
, {Zhang}, X\BPBI J.%
, {Demekhov}, A\BPBI G.%
, {Meng}, X.%
, {Angelopoulos}, V.%
\BCBL {}\ \BBA {} {Fedorenko}, Y\BPBI V.%
\end{APACrefauthors}%
\unskip\
\newblock
\APACrefYearMonthDay{2024}{{\APACmonth{02}}}{}.
\newblock
{\BBOQ}\APACrefatitle {{Relativistic Electron Precipitation Driven by Mesoscale Transients, Inferred From Ground and Multi-Spacecraft Platforms}} {{Relativistic Electron Precipitation Driven by Mesoscale Transients, Inferred From Ground and Multi-Spacecraft Platforms}}.{\BBCQ}
\newblock
\APACjournalVolNumPages{Journal of Geophysical Research (Space Physics)}{129}{2}{e2023JA032287}.
\newblock
\begin{APACrefDOI} \doi{10.1029/2023JA032287} \end{APACrefDOI}
\PrintBackRefs{\CurrentBib}

\bibitem [\protect \citeauthoryear {%
{Baker}%
, {Pulkkinen}%
, {Angelopoulos}%
, {Baumjohann}%
\BCBL {}\ \BBA {} {McPherron}%
}{%
{Baker}%
\ \protect \BOthers {.}}{%
{\protect \APACyear {1996}}%
}]{%
Baker96}
\APACinsertmetastar {%
Baker96}%
\begin{APACrefauthors}%
{Baker}, D\BPBI N.%
, {Pulkkinen}, T\BPBI I.%
, {Angelopoulos}, V.%
, {Baumjohann}, W.%
\BCBL {}\ \BBA {} {McPherron}, R\BPBI L.%
\end{APACrefauthors}%
\unskip\
\newblock
\APACrefYearMonthDay{1996}{{\APACmonth{06}}}{}.
\newblock
{\BBOQ}\APACrefatitle {{Neutral line model of substorms: Past results and present view}} {{Neutral line model of substorms: Past results and present view}}.{\BBCQ}
\newblock
\APACjournalVolNumPages{\jgr}{101}{}{12975-13010}.
\newblock
\begin{APACrefDOI} \doi{10.1029/95JA03753} \end{APACrefDOI}
\PrintBackRefs{\CurrentBib}

\bibitem [\protect \citeauthoryear {%
{Birmingham}%
}{%
{Birmingham}%
}{%
{\protect \APACyear {1984}}%
}]{%
Birmingham84}
\APACinsertmetastar {%
Birmingham84}%
\begin{APACrefauthors}%
{Birmingham}, T\BPBI J.%
\end{APACrefauthors}%
\unskip\
\newblock
\APACrefYearMonthDay{1984}{{\APACmonth{05}}}{}.
\newblock
{\BBOQ}\APACrefatitle {{Pitch angle diffusion in the Jovian magnetodisc}} {{Pitch angle diffusion in the Jovian magnetodisc}}.{\BBCQ}
\newblock
\APACjournalVolNumPages{\jgr}{89}{}{2699-2707}.
\newblock
\begin{APACrefDOI} \doi{10.1029/JA089iA05p02699} \end{APACrefDOI}
\PrintBackRefs{\CurrentBib}

\bibitem [\protect \citeauthoryear {%
{Birn}%
, {Merkin}%
, {Sitnov}%
\BCBL {}\ \BBA {} {Otto}%
}{%
{Birn}%
\ \protect \BOthers {.}}{%
{\protect \APACyear {2018}}%
}]{%
Birn18}
\APACinsertmetastar {%
Birn18}%
\begin{APACrefauthors}%
{Birn}, J.%
, {Merkin}, V\BPBI G.%
, {Sitnov}, M\BPBI I.%
\BCBL {}\ \BBA {} {Otto}, A.%
\end{APACrefauthors}%
\unskip\
\newblock
\APACrefYearMonthDay{2018}{{\APACmonth{05}}}{}.
\newblock
{\BBOQ}\APACrefatitle {{MHD Stability of Magnetotail Configurations With a $B_{z}$ Hump}} {{MHD Stability of Magnetotail Configurations With a $B_{z}$ Hump}}.{\BBCQ}
\newblock
\APACjournalVolNumPages{\jgr}{123}{}{3477-3492}.
\newblock
\begin{APACrefDOI} \doi{10.1029/2018JA025290} \end{APACrefDOI}
\PrintBackRefs{\CurrentBib}

\bibitem [\protect \citeauthoryear {%
{Birn}%
, {Runov}%
\BCBL {}\ \BBA {} {Khotyaintsev}%
}{%
{Birn}%
\ \protect \BOthers {.}}{%
{\protect \APACyear {2021}}%
}]{%
Birn21:AGU}
\APACinsertmetastar {%
Birn21:AGU}%
\begin{APACrefauthors}%
{Birn}, J.%
, {Runov}, A.%
\BCBL {}\ \BBA {} {Khotyaintsev}, Y.%
\end{APACrefauthors}%
\unskip\
\newblock
\APACrefYearMonthDay{2021}{{\APACmonth{05}}}{}.
\newblock
{\BBOQ}\APACrefatitle {{Magnetotail Processes}} {{Magnetotail Processes}}.{\BBCQ}
\newblock
\BIn{} R.~{Maggiolo}, N.~{Andr{\'e}}, H.~{Hasegawa}\BCBL {}\ \BBA {} D\BPBI T.~{Welling}\ (\BEDS), \APACrefbtitle {Magnetospheres in the Solar System} {Magnetospheres in the solar system}\ (\BVOL~2, \BPG~245).
\newblock
\begin{APACrefDOI} \doi{10.1002/9781119815624.ch17} \end{APACrefDOI}
\PrintBackRefs{\CurrentBib}

\bibitem [\protect \citeauthoryear {%
{Blake}%
\ \protect \BOthers {.}}{%
{Blake}%
\ \protect \BOthers {.}}{%
{\protect \APACyear {2016}}%
}]{%
Blake16}
\APACinsertmetastar {%
Blake16}%
\begin{APACrefauthors}%
{Blake}, J\BPBI B.%
, {Mauk}, B\BPBI H.%
, {Baker}, D\BPBI N.%
, {Carranza}, P.%
, {Clemmons}, J\BPBI H.%
, {Craft}, J.%
\BDBL {}{Westlake}, J.%
\end{APACrefauthors}%
\unskip\
\newblock
\APACrefYearMonthDay{2016}{{\APACmonth{03}}}{}.
\newblock
{\BBOQ}\APACrefatitle {{The Fly's Eye Energetic Particle Spectrometer (FEEPS) Sensors for the Magnetospheric Multiscale (MMS) Mission}} {{The Fly's Eye Energetic Particle Spectrometer (FEEPS) Sensors for the Magnetospheric Multiscale (MMS) Mission}}.{\BBCQ}
\newblock
\APACjournalVolNumPages{\ssr}{199}{}{309-329}.
\newblock
\begin{APACrefDOI} \doi{10.1007/s11214-015-0163-x} \end{APACrefDOI}
\PrintBackRefs{\CurrentBib}

\bibitem [\protect \citeauthoryear {%
Bortnik%
, Li%
, Thorne%
\BCBL {}\ \BBA {} Angelopoulos%
}{%
Bortnik%
\ \protect \BOthers {.}}{%
{\protect \APACyear {2016}}%
}]{%
Bortnik16}
\APACinsertmetastar {%
Bortnik16}%
\begin{APACrefauthors}%
Bortnik, J.%
, Li, W.%
, Thorne, R\BPBI M.%
\BCBL {}\ \BBA {} Angelopoulos, V.%
\end{APACrefauthors}%
\unskip\
\newblock
\APACrefYearMonthDay{2016}{}{}.
\newblock
{\BBOQ}\APACrefatitle {A unified approach to inner magnetospheric state prediction} {A unified approach to inner magnetospheric state prediction}.{\BBCQ}
\newblock
\APACjournalVolNumPages{Journal of Geophysical Research: Space Physics}{121}{3}{2423-2430}.
\newblock
\begin{APACrefURL} \url{https://agupubs.onlinelibrary.wiley.com/doi/abs/10.1002/2015JA021733} \end{APACrefURL}
\newblock
\begin{APACrefDOI} \doi{https://doi.org/10.1002/2015JA021733} \end{APACrefDOI}
\PrintBackRefs{\CurrentBib}

\bibitem [\protect \citeauthoryear {%
{Burch}%
, {Moore}%
, {Torbert}%
\BCBL {}\ \BBA {} {Giles}%
}{%
{Burch}%
\ \protect \BOthers {.}}{%
{\protect \APACyear {2016}}%
}]{%
Burch16}
\APACinsertmetastar {%
Burch16}%
\begin{APACrefauthors}%
{Burch}, J\BPBI L.%
, {Moore}, T\BPBI E.%
, {Torbert}, R\BPBI B.%
\BCBL {}\ \BBA {} {Giles}, B\BPBI L.%
\end{APACrefauthors}%
\unskip\
\newblock
\APACrefYearMonthDay{2016}{{\APACmonth{03}}}{}.
\newblock
{\BBOQ}\APACrefatitle {{Magnetospheric Multiscale Overview and Science Objectives}} {{Magnetospheric Multiscale Overview and Science Objectives}}.{\BBCQ}
\newblock
\APACjournalVolNumPages{\ssr}{199}{}{5-21}.
\newblock
\begin{APACrefDOI} \doi{10.1007/s11214-015-0164-9} \end{APACrefDOI}
\PrintBackRefs{\CurrentBib}

\bibitem [\protect \citeauthoryear {%
Coxon%
, Milan%
\BCBL {}\ \BBA {} Anderson%
}{%
Coxon%
\ \protect \BOthers {.}}{%
{\protect \APACyear {2018}}%
}]{%
Coxon18}
\APACinsertmetastar {%
Coxon18}%
\begin{APACrefauthors}%
Coxon, J\BPBI C.%
, Milan, S\BPBI E.%
\BCBL {}\ \BBA {} Anderson, B\BPBI J.%
\end{APACrefauthors}%
\unskip\
\newblock
\APACrefYearMonthDay{2018}{}{}.
\newblock
{\BBOQ}\APACrefatitle {A Review of Birkeland Current Research Using AMPERE} {A review of birkeland current research using ampere}.{\BBCQ}
\newblock
\BIn{} \APACrefbtitle {Electric Currents in Geospace and Beyond} {Electric currents in geospace and beyond}\ (\BPG~257-278).
\newblock
\APACaddressPublisher{}{American Geophysical Union (AGU)}.
\newblock
\begin{APACrefURL} \url{https://agupubs.onlinelibrary.wiley.com/doi/abs/10.1002/9781119324522.ch16} \end{APACrefURL}
\newblock
\begin{APACrefDOI} \doi{https://doi.org/10.1002/9781119324522.ch16} \end{APACrefDOI}
\PrintBackRefs{\CurrentBib}

\bibitem [\protect \citeauthoryear {%
{Delcourt}%
, {Martin}%
\BCBL {}\ \BBA {} {Alem}%
}{%
{Delcourt}%
\ \protect \BOthers {.}}{%
{\protect \APACyear {1994}}%
}]{%
Delcourt94:scattering}
\APACinsertmetastar {%
Delcourt94:scattering}%
\begin{APACrefauthors}%
{Delcourt}, D\BPBI C.%
, {Martin}, R\BPBI F., Jr.%
\BCBL {}\ \BBA {} {Alem}, F.%
\end{APACrefauthors}%
\unskip\
\newblock
\APACrefYearMonthDay{1994}{{\APACmonth{07}}}{}.
\newblock
{\BBOQ}\APACrefatitle {{A simple model of magnetic moment scattering in a field reversal}} {{A simple model of magnetic moment scattering in a field reversal}}.{\BBCQ}
\newblock
\APACjournalVolNumPages{\grl}{21}{}{1543-1546}.
\newblock
\begin{APACrefDOI} \doi{10.1029/94GL01291} \end{APACrefDOI}
\PrintBackRefs{\CurrentBib}

\bibitem [\protect \citeauthoryear {%
{Dubyagin}%
, {Sergeev}%
\BCBL {}\ \BBA {} {Kubyshkina}%
}{%
{Dubyagin}%
\ \protect \BOthers {.}}{%
{\protect \APACyear {2002}}%
}]{%
Dubyagin02}
\APACinsertmetastar {%
Dubyagin02}%
\begin{APACrefauthors}%
{Dubyagin}, S.%
, {Sergeev}, V\BPBI A.%
\BCBL {}\ \BBA {} {Kubyshkina}, M\BPBI V.%
\end{APACrefauthors}%
\unskip\
\newblock
\APACrefYearMonthDay{2002}{{\APACmonth{03}}}{}.
\newblock
{\BBOQ}\APACrefatitle {{On the remote sensing of plasma sheet from low-altitude spacecraft}} {{On the remote sensing of plasma sheet from low-altitude spacecraft}}.{\BBCQ}
\newblock
\APACjournalVolNumPages{Journal of Atmospheric and Solar-Terrestrial Physics}{64}{5-6}{567-572}.
\newblock
\begin{APACrefDOI} \doi{10.1016/S1364-6826(02)00014-7} \end{APACrefDOI}
\PrintBackRefs{\CurrentBib}

\bibitem [\protect \citeauthoryear {%
{Erkaev}%
, {Semenov}%
\BCBL {}\ \BBA {} {Biernat}%
}{%
{Erkaev}%
\ \protect \BOthers {.}}{%
{\protect \APACyear {2007}}%
}]{%
Erkaev07}
\APACinsertmetastar {%
Erkaev07}%
\begin{APACrefauthors}%
{Erkaev}, N\BPBI V.%
, {Semenov}, V\BPBI S.%
\BCBL {}\ \BBA {} {Biernat}, H\BPBI K.%
\end{APACrefauthors}%
\unskip\
\newblock
\APACrefYearMonthDay{2007}{{\APACmonth{12}}}{}.
\newblock
{\BBOQ}\APACrefatitle {{Magnetic Double-Gradient Instability and Flapping Waves in a Current Sheet}} {{Magnetic Double-Gradient Instability and Flapping Waves in a Current Sheet}}.{\BBCQ}
\newblock
\APACjournalVolNumPages{Physical Review Letters}{99}{23}{235003}.
\newblock
\begin{APACrefDOI} \doi{10.1103/PhysRevLett.99.235003} \end{APACrefDOI}
\PrintBackRefs{\CurrentBib}

\bibitem [\protect \citeauthoryear {%
{Gjerloev}%
}{%
{Gjerloev}%
}{%
{\protect \APACyear {2009}}%
}]{%
Gjerloev09}
\APACinsertmetastar {%
Gjerloev09}%
\begin{APACrefauthors}%
{Gjerloev}, J\BPBI W.%
\end{APACrefauthors}%
\unskip\
\newblock
\APACrefYearMonthDay{2009}{{\APACmonth{07}}}{}.
\newblock
{\BBOQ}\APACrefatitle {{A Global Ground-Based Magnetometer Initiative}} {{A Global Ground-Based Magnetometer Initiative}}.{\BBCQ}
\newblock
\APACjournalVolNumPages{EOS Transactions}{90}{27}{230-231}.
\newblock
\begin{APACrefDOI} \doi{10.1029/2009EO270002} \end{APACrefDOI}
\PrintBackRefs{\CurrentBib}

\bibitem [\protect \citeauthoryear {%
Gjerloev%
}{%
Gjerloev%
}{%
{\protect \APACyear {2012}}%
}]{%
Gjerloev2012}
\APACinsertmetastar {%
Gjerloev2012}%
\begin{APACrefauthors}%
Gjerloev, J\BPBI W.%
\end{APACrefauthors}%
\unskip\
\newblock
\APACrefYearMonthDay{2012}{}{}.
\newblock
{\BBOQ}\APACrefatitle {The SuperMAG data processing technique} {The supermag data processing technique}.{\BBCQ}
\newblock
\APACjournalVolNumPages{Journal of Geophysical Research: Space Physics}{117}{A9}{https://doi.org/10.1029/2012JA017683}.
\newblock
\begin{APACrefDOI} \doi{https://doi.org/10.1029/2012JA017683} \end{APACrefDOI}
\PrintBackRefs{\CurrentBib}

\bibitem [\protect \citeauthoryear {%
{Gkioulidou}%
\ \protect \BOthers {.}}{%
{Gkioulidou}%
\ \protect \BOthers {.}}{%
{\protect \APACyear {2014}}%
}]{%
Gkioulidou14}
\APACinsertmetastar {%
Gkioulidou14}%
\begin{APACrefauthors}%
{Gkioulidou}, M.%
, {Ukhorskiy}, A\BPBI Y.%
, {Mitchell}, D\BPBI G.%
, {Sotirelis}, T.%
, {Mauk}, B\BPBI H.%
\BCBL {}\ \BBA {} {Lanzerotti}, L\BPBI J.%
\end{APACrefauthors}%
\unskip\
\newblock
\APACrefYearMonthDay{2014}{{\APACmonth{09}}}{}.
\newblock
{\BBOQ}\APACrefatitle {{The role of small-scale ion injections in the buildup of Earth's ring current pressure: Van Allen Probes observations of the 17 March 2013 storm}} {{The role of small-scale ion injections in the buildup of Earth's ring current pressure: Van Allen Probes observations of the 17 March 2013 storm}}.{\BBCQ}
\newblock
\APACjournalVolNumPages{\jgr}{119}{}{7327-7342}.
\newblock
\begin{APACrefDOI} \doi{10.1002/2014JA020096} \end{APACrefDOI}
\PrintBackRefs{\CurrentBib}

\bibitem [\protect \citeauthoryear {%
{Gonzalez}%
\ \BBA {} {Parker}%
}{%
{Gonzalez}%
\ \BBA {} {Parker}%
}{%
{\protect \APACyear {2016}}%
}]{%
book:Gonzalez&Parker}
\APACinsertmetastar {%
book:Gonzalez&Parker}%
\begin{APACrefauthors}%
{Gonzalez}, W.%
\BCBT {}\ \BBA {} {Parker}, E.%
\end{APACrefauthors}%
\unskip\
\newblock
\APACrefYear{2016}.
\newblock
\APACrefbtitle {{Magnetic Reconnection}} {{Magnetic Reconnection}}\ (\BVOL~427).
\newblock
\begin{APACrefDOI} \doi{10.1007/978-3-319-26432-5} \end{APACrefDOI}
\PrintBackRefs{\CurrentBib}

\bibitem [\protect \citeauthoryear {%
{Imhof}%
, {Reagan}%
\BCBL {}\ \BBA {} {Gaines}%
}{%
{Imhof}%
\ \protect \BOthers {.}}{%
{\protect \APACyear {1977}}%
}]{%
Imhof77}
\APACinsertmetastar {%
Imhof77}%
\begin{APACrefauthors}%
{Imhof}, W\BPBI L.%
, {Reagan}, J\BPBI B.%
\BCBL {}\ \BBA {} {Gaines}, E\BPBI E.%
\end{APACrefauthors}%
\unskip\
\newblock
\APACrefYearMonthDay{1977}{{\APACmonth{11}}}{}.
\newblock
{\BBOQ}\APACrefatitle {{Fine-scale spatial structure in the pitch angle distributions of energetic particles near the midnight trapping boundary}} {{Fine-scale spatial structure in the pitch angle distributions of energetic particles near the midnight trapping boundary}}.{\BBCQ}
\newblock
\APACjournalVolNumPages{\jgr}{82}{}{5215-5221}.
\newblock
\begin{APACrefDOI} \doi{10.1029/JA082i032p05215} \end{APACrefDOI}
\PrintBackRefs{\CurrentBib}

\bibitem [\protect \citeauthoryear {%
{Kozelova}%
\ \BBA {} {Kozelov}%
}{%
{Kozelova}%
\ \BBA {} {Kozelov}%
}{%
{\protect \APACyear {2013}}%
}]{%
Kozelova&Kozelov13}
\APACinsertmetastar {%
Kozelova&Kozelov13}%
\begin{APACrefauthors}%
{Kozelova}, T\BPBI V.%
\BCBT {}\ \BBA {} {Kozelov}, B\BPBI V.%
\end{APACrefauthors}%
\unskip\
\newblock
\APACrefYearMonthDay{2013}{{\APACmonth{06}}}{}.
\newblock
{\BBOQ}\APACrefatitle {{Substorm-associated explosive magnetic field stretching near the earthward edge of the plasma sheet}} {{Substorm-associated explosive magnetic field stretching near the earthward edge of the plasma sheet}}.{\BBCQ}
\newblock
\APACjournalVolNumPages{Journal of Geophysical Research (Space Physics)}{118}{6}{3323-3335}.
\newblock
\begin{APACrefDOI} \doi{10.1002/jgra.50344} \end{APACrefDOI}
\PrintBackRefs{\CurrentBib}

\bibitem [\protect \citeauthoryear {%
{Kubyshkina}%
\ \protect \BOthers {.}}{%
{Kubyshkina}%
\ \protect \BOthers {.}}{%
{\protect \APACyear {2011}}%
}]{%
Kubyshkina11}
\APACinsertmetastar {%
Kubyshkina11}%
\begin{APACrefauthors}%
{Kubyshkina}, M.%
, {Sergeev}, V.%
, {Tsyganenko}, N.%
, {Angelopoulos}, V.%
, {Runov}, A.%
, {Donovan}, E.%
\BDBL {}{Baumjohann}, W.%
\end{APACrefauthors}%
\unskip\
\newblock
\APACrefYearMonthDay{2011}{{\APACmonth{02}}}{}.
\newblock
{\BBOQ}\APACrefatitle {{Time-dependent magnetospheric configuration and breakup mapping during a substorm}} {{Time-dependent magnetospheric configuration and breakup mapping during a substorm}}.{\BBCQ}
\newblock
\APACjournalVolNumPages{\jgr}{116}{}{0}.
\newblock
\begin{APACrefDOI} \doi{10.1029/2010JA015882} \end{APACrefDOI}
\PrintBackRefs{\CurrentBib}

\bibitem [\protect \citeauthoryear {%
{Kubyshkina}%
\ \protect \BOthers {.}}{%
{Kubyshkina}%
\ \protect \BOthers {.}}{%
{\protect \APACyear {2009}}%
}]{%
Kubyshkina09}
\APACinsertmetastar {%
Kubyshkina09}%
\begin{APACrefauthors}%
{Kubyshkina}, M.%
, {Sergeev}, V.%
, {Tsyganenko}, N.%
, {Angelopoulos}, V.%
, {Runov}, A.%
, {Singer}, H.%
\BDBL {}{Baumjohann}, W.%
\end{APACrefauthors}%
\unskip\
\newblock
\APACrefYearMonthDay{2009}{{\APACmonth{04}}}{}.
\newblock
{\BBOQ}\APACrefatitle {{Toward adapted time-dependent magnetospheric models: A simple approach based on tuning the standard model}} {{Toward adapted time-dependent magnetospheric models: A simple approach based on tuning the standard model}}.{\BBCQ}
\newblock
\APACjournalVolNumPages{\jgr}{114}{}{0}.
\newblock
\begin{APACrefDOI} \doi{10.1029/2008JA013547} \end{APACrefDOI}
\PrintBackRefs{\CurrentBib}

\bibitem [\protect \citeauthoryear {%
{Lin}%
, {Wang}%
, {Lu}%
, {Perez}%
\BCBL {}\ \BBA {} {Lu}%
}{%
{Lin}%
\ \protect \BOthers {.}}{%
{\protect \APACyear {2014}}%
}]{%
Lin14:hybrid_code}
\APACinsertmetastar {%
Lin14:hybrid_code}%
\begin{APACrefauthors}%
{Lin}, Y.%
, {Wang}, X\BPBI Y.%
, {Lu}, S.%
, {Perez}, J\BPBI D.%
\BCBL {}\ \BBA {} {Lu}, Q.%
\end{APACrefauthors}%
\unskip\
\newblock
\APACrefYearMonthDay{2014}{{\APACmonth{09}}}{}.
\newblock
{\BBOQ}\APACrefatitle {{Investigation of storm time magnetotail and ion injection using three-dimensional global hybrid simulation}} {{Investigation of storm time magnetotail and ion injection using three-dimensional global hybrid simulation}}.{\BBCQ}
\newblock
\APACjournalVolNumPages{\jgr}{119}{}{7413-7432}.
\newblock
\begin{APACrefDOI} \doi{10.1002/2014JA020005} \end{APACrefDOI}
\PrintBackRefs{\CurrentBib}

\bibitem [\protect \citeauthoryear {%
Menk%
\ \BBA {} Waters%
}{%
Menk%
\ \BBA {} Waters%
}{%
{\protect \APACyear {2013}}%
}]{%
MenkWaters13}
\APACinsertmetastar {%
MenkWaters13}%
\begin{APACrefauthors}%
Menk, F\BPBI W.%
\BCBT {}\ \BBA {} Waters, C\BPBI L.%
\end{APACrefauthors}%
\unskip\
\newblock
\APACrefYear{2013}.
\newblock
\APACrefbtitle {Magnetoseismology: Ground-based remote sensing of Earth's magnetosphere} {Magnetoseismology: Ground-based remote sensing of earth's magnetosphere}.
\newblock
\APACaddressPublisher{}{John Wiley \& Sons}.
\PrintBackRefs{\CurrentBib}

\bibitem [\protect \citeauthoryear {%
{Mourenas}%
\ \protect \BOthers {.}}{%
{Mourenas}%
\ \protect \BOthers {.}}{%
{\protect \APACyear {2021}}%
}]{%
Mourenas21:jgr:ELFIN}
\APACinsertmetastar {%
Mourenas21:jgr:ELFIN}%
\begin{APACrefauthors}%
{Mourenas}, D.%
, {Artemyev}, A\BPBI V.%
, {Zhang}, X\BPBI J.%
, {Angelopoulos}, V.%
, {Tsai}, E.%
\BCBL {}\ \BBA {} {Wilkins}, C.%
\end{APACrefauthors}%
\unskip\
\newblock
\APACrefYearMonthDay{2021}{{\APACmonth{11}}}{}.
\newblock
{\BBOQ}\APACrefatitle {{Electron Lifetimes and Diffusion Rates Inferred From ELFIN Measurements at Low Altitude: First Results}} {{Electron Lifetimes and Diffusion Rates Inferred From ELFIN Measurements at Low Altitude: First Results}}.{\BBCQ}
\newblock
\APACjournalVolNumPages{Journal of Geophysical Research (Space Physics)}{126}{11}{e29757}.
\newblock
\begin{APACrefDOI} \doi{10.1029/2021JA029757} \end{APACrefDOI}
\PrintBackRefs{\CurrentBib}

\bibitem [\protect \citeauthoryear {%
{Panov}%
\ \BBA {} {Pritchett}%
}{%
{Panov}%
\ \BBA {} {Pritchett}%
}{%
{\protect \APACyear {2018}}%
}]{%
Panov&Pritchett18}
\APACinsertmetastar {%
Panov&Pritchett18}%
\begin{APACrefauthors}%
{Panov}, E\BPBI V.%
\BCBT {}\ \BBA {} {Pritchett}, P\BPBI L.%
\end{APACrefauthors}%
\unskip\
\newblock
\APACrefYearMonthDay{2018}{{\APACmonth{09}}}{}.
\newblock
{\BBOQ}\APACrefatitle {{Dawnward Drifting Interchange Heads in the Earth's Magnetotail}} {{Dawnward Drifting Interchange Heads in the Earth's Magnetotail}}.{\BBCQ}
\newblock
\APACjournalVolNumPages{\grl}{45}{17}{8834-8843}.
\newblock
\begin{APACrefDOI} \doi{10.1029/2018GL078482} \end{APACrefDOI}
\PrintBackRefs{\CurrentBib}

\bibitem [\protect \citeauthoryear {%
{Petrukovich}%
\ \protect \BOthers {.}}{%
{Petrukovich}%
\ \protect \BOthers {.}}{%
{\protect \APACyear {2007}}%
}]{%
Petrukovich07}
\APACinsertmetastar {%
Petrukovich07}%
\begin{APACrefauthors}%
{Petrukovich}, A\BPBI A.%
, {Baumjohann}, W.%
, {Nakamura}, R.%
, {Runov}, A.%
, {Balogh}, A.%
\BCBL {}\ \BBA {} {R{\`e}me}, H.%
\end{APACrefauthors}%
\unskip\
\newblock
\APACrefYearMonthDay{2007}{{\APACmonth{10}}}{}.
\newblock
{\BBOQ}\APACrefatitle {{Thinning and stretching of the plasma sheet}} {{Thinning and stretching of the plasma sheet}}.{\BBCQ}
\newblock
\APACjournalVolNumPages{\jgr}{112}{}{10213}.
\newblock
\begin{APACrefDOI} \doi{10.1029/2007JA012349} \end{APACrefDOI}
\PrintBackRefs{\CurrentBib}

\bibitem [\protect \citeauthoryear {%
Pritchett%
\ \BBA {} Coroniti%
}{%
Pritchett%
\ \BBA {} Coroniti%
}{%
{\protect \APACyear {2010}}%
}]{%
Pritchett&Coroniti2010}
\APACinsertmetastar {%
Pritchett&Coroniti2010}%
\begin{APACrefauthors}%
Pritchett, P\BPBI L.%
\BCBT {}\ \BBA {} Coroniti, F\BPBI V.%
\end{APACrefauthors}%
\unskip\
\newblock
\APACrefYearMonthDay{2010}{}{}.
\newblock
{\BBOQ}\APACrefatitle {A kinetic ballooning/interchange instability in the magnetotail} {A kinetic ballooning/interchange instability in the magnetotail}.{\BBCQ}
\newblock
\APACjournalVolNumPages{Journal of Geophysical Research: Space Physics}{115}{A6}{}.
\newblock
\begin{APACrefDOI} \doi{https://doi.org/10.1029/2009JA014752} \end{APACrefDOI}
\PrintBackRefs{\CurrentBib}

\bibitem [\protect \citeauthoryear {%
{Runov}%
\ \protect \BOthers {.}}{%
{Runov}%
\ \protect \BOthers {.}}{%
{\protect \APACyear {2021}}%
}]{%
Runov21:jastp}
\APACinsertmetastar {%
Runov21:jastp}%
\begin{APACrefauthors}%
{Runov}, A.%
, {Angelopoulos}, V.%
, {Artemyev}, A\BPBI V.%
, {Weygand}, J\BPBI M.%
, {Lu}, S.%
, {Lin}, Y.%
\BCBL {}\ \BBA {} {Zhang}, X\BPBI J.%
\end{APACrefauthors}%
\unskip\
\newblock
\APACrefYearMonthDay{2021}{{\APACmonth{09}}}{}.
\newblock
{\BBOQ}\APACrefatitle {{Global and local processes of thin current sheet formation during substorm growth phase}} {{Global and local processes of thin current sheet formation during substorm growth phase}}.{\BBCQ}
\newblock
\APACjournalVolNumPages{Journal of Atmospheric and Solar-Terrestrial Physics}{220}{}{105671}.
\newblock
\begin{APACrefDOI} \doi{10.1016/j.jastp.2021.105671} \end{APACrefDOI}
\PrintBackRefs{\CurrentBib}

\bibitem [\protect \citeauthoryear {%
{Sergeev}%
\ \protect \BOthers {.}}{%
{Sergeev}%
\ \protect \BOthers {.}}{%
{\protect \APACyear {2011}}%
}]{%
Sergeev11}
\APACinsertmetastar {%
Sergeev11}%
\begin{APACrefauthors}%
{Sergeev}, V\BPBI A.%
, {Angelopoulos}, V.%
, {Kubyshkina}, M.%
, {Donovan}, E.%
, {Zhou}, X\BHBI Z.%
, {Runov}, A.%
\BDBL {}{Nakamura}, R.%
\end{APACrefauthors}%
\unskip\
\newblock
\APACrefYearMonthDay{2011}{{\APACmonth{02}}}{}.
\newblock
{\BBOQ}\APACrefatitle {{Substorm growth and expansion onset as observed with ideal ground-spacecraft THEMIS coverage}} {{Substorm growth and expansion onset as observed with ideal ground-spacecraft THEMIS coverage}}.{\BBCQ}
\newblock
\APACjournalVolNumPages{\jgr}{116}{}{A00I26}.
\newblock
\begin{APACrefDOI} \doi{10.1029/2010JA015689} \end{APACrefDOI}
\PrintBackRefs{\CurrentBib}

\bibitem [\protect \citeauthoryear {%
{Sergeev}%
, {Angelopoulos}%
\BCBL {}\ \BBA {} {Nakamura}%
}{%
{Sergeev}%
, {Angelopoulos}%
\BCBL {}\ \BBA {} {Nakamura}%
}{%
{\protect \APACyear {2012}}%
}]{%
Sergeev12}
\APACinsertmetastar {%
Sergeev12}%
\begin{APACrefauthors}%
{Sergeev}, V\BPBI A.%
, {Angelopoulos}, V.%
\BCBL {}\ \BBA {} {Nakamura}, R.%
\end{APACrefauthors}%
\unskip\
\newblock
\APACrefYearMonthDay{2012}{{\APACmonth{03}}}{}.
\newblock
{\BBOQ}\APACrefatitle {{Recent advances in understanding substorm dynamics}} {{Recent advances in understanding substorm dynamics}}.{\BBCQ}
\newblock
\APACjournalVolNumPages{\grl}{39}{}{5101}.
\newblock
\begin{APACrefDOI} \doi{10.1029/2012GL050859} \end{APACrefDOI}
\PrintBackRefs{\CurrentBib}

\bibitem [\protect \citeauthoryear {%
{Sergeev}%
, {Chernyaev}%
, {Angelopoulos}%
\BCBL {}\ \BBA {} {Ganushkina}%
}{%
{Sergeev}%
\ \protect \BOthers {.}}{%
{\protect \APACyear {2015}}%
}]{%
Sergeev15}
\APACinsertmetastar {%
Sergeev15}%
\begin{APACrefauthors}%
{Sergeev}, V\BPBI A.%
, {Chernyaev}, I\BPBI A.%
, {Angelopoulos}, V.%
\BCBL {}\ \BBA {} {Ganushkina}, N\BPBI Y.%
\end{APACrefauthors}%
\unskip\
\newblock
\APACrefYearMonthDay{2015}{{\APACmonth{12}}}{}.
\newblock
{\BBOQ}\APACrefatitle {{Magnetospheric conditions near the equatorial footpoints of proton isotropy boundaries}} {{Magnetospheric conditions near the equatorial footpoints of proton isotropy boundaries}}.{\BBCQ}
\newblock
\APACjournalVolNumPages{Annales Geophysicae}{33}{}{1485-1493}.
\newblock
\begin{APACrefDOI} \doi{10.5194/angeo-33-1485-2015} \end{APACrefDOI}
\PrintBackRefs{\CurrentBib}

\bibitem [\protect \citeauthoryear {%
{Sergeev}%
, {Gordeev}%
, {Merkin}%
\BCBL {}\ \BBA {} {Sitnov}%
}{%
{Sergeev}%
\ \protect \BOthers {.}}{%
{\protect \APACyear {2018}}%
}]{%
Sergeev18:grl}
\APACinsertmetastar {%
Sergeev18:grl}%
\begin{APACrefauthors}%
{Sergeev}, V\BPBI A.%
, {Gordeev}, E\BPBI I.%
, {Merkin}, V\BPBI G.%
\BCBL {}\ \BBA {} {Sitnov}, M\BPBI I.%
\end{APACrefauthors}%
\unskip\
\newblock
\APACrefYearMonthDay{2018}{{\APACmonth{03}}}{}.
\newblock
{\BBOQ}\APACrefatitle {{Does a Local B-Minimum Appear in the Tail Current Sheet During a Substorm Growth Phase?}} {{Does a Local B-Minimum Appear in the Tail Current Sheet During a Substorm Growth Phase?}}{\BBCQ}
\newblock
\APACjournalVolNumPages{\grl}{45}{}{2566-2573}.
\newblock
\begin{APACrefDOI} \doi{10.1002/2018GL077183} \end{APACrefDOI}
\PrintBackRefs{\CurrentBib}

\bibitem [\protect \citeauthoryear {%
{Sergeev}%
\ \protect \BOthers {.}}{%
{Sergeev}%
\ \protect \BOthers {.}}{%
{\protect \APACyear {2023}}%
}]{%
Sergeev23:elfin}
\APACinsertmetastar {%
Sergeev23:elfin}%
\begin{APACrefauthors}%
{Sergeev}, V\BPBI A.%
, {Kubyshkina}, M\BPBI V.%
, {Semenov}, V\BPBI S.%
, {Artemyev}, A.%
, {Angelopoulos}, V.%
\BCBL {}\ \BBA {} {Runov}, A.%
\end{APACrefauthors}%
\unskip\
\newblock
\APACrefYearMonthDay{2023}{{\APACmonth{11}}}{}.
\newblock
{\BBOQ}\APACrefatitle {{Unusual Magnetospheric Dynamics During Intense Substorm Initiated by Strong Magnetospheric Compression}} {{Unusual Magnetospheric Dynamics During Intense Substorm Initiated by Strong Magnetospheric Compression}}.{\BBCQ}
\newblock
\APACjournalVolNumPages{Journal of Geophysical Research (Space Physics)}{128}{11}{e2023JA031536}.
\newblock
\begin{APACrefDOI} \doi{10.1029/2023JA031536} \end{APACrefDOI}
\PrintBackRefs{\CurrentBib}

\bibitem [\protect \citeauthoryear {%
{Sergeev}%
, {Nishimura}%
\BCBL {}\ \protect \BOthers {.}}{%
{Sergeev}%
, {Nishimura}%
\BCBL {}\ \protect \BOthers {.}}{%
{\protect \APACyear {2012}}%
}]{%
Sergeev12:IB}
\APACinsertmetastar {%
Sergeev12:IB}%
\begin{APACrefauthors}%
{Sergeev}, V\BPBI A.%
, {Nishimura}, Y.%
, {Kubyshkina}, M.%
, {Angelopoulos}, V.%
, {Nakamura}, R.%
\BCBL {}\ \BBA {} {Singer}, H.%
\end{APACrefauthors}%
\unskip\
\newblock
\APACrefYearMonthDay{2012}{{\APACmonth{01}}}{}.
\newblock
{\BBOQ}\APACrefatitle {{Magnetospheric location of the equatorward prebreakup arc}} {{Magnetospheric location of the equatorward prebreakup arc}}.{\BBCQ}
\newblock
\APACjournalVolNumPages{Journal of Geophysical Research (Space Physics)}{117}{A1}{A01212}.
\newblock
\begin{APACrefDOI} \doi{10.1029/2011JA017154} \end{APACrefDOI}
\PrintBackRefs{\CurrentBib}

\bibitem [\protect \citeauthoryear {%
{Sergeev}%
\ \BBA {} {Tsyganenko}%
}{%
{Sergeev}%
\ \BBA {} {Tsyganenko}%
}{%
{\protect \APACyear {1982}}%
}]{%
Sergeev&Tsyganenko82}
\APACinsertmetastar {%
Sergeev&Tsyganenko82}%
\begin{APACrefauthors}%
{Sergeev}, V\BPBI A.%
\BCBT {}\ \BBA {} {Tsyganenko}, N\BPBI A.%
\end{APACrefauthors}%
\unskip\
\newblock
\APACrefYearMonthDay{1982}{{\APACmonth{10}}}{}.
\newblock
{\BBOQ}\APACrefatitle {{Energetic particle losses and trapping boundaries as deduced from calculations with a realistic magnetic field model}} {{Energetic particle losses and trapping boundaries as deduced from calculations with a realistic magnetic field model}}.{\BBCQ}
\newblock
\APACjournalVolNumPages{\planss}{30}{}{999-1006}.
\newblock
\begin{APACrefDOI} \doi{10.1016/0032-0633(82)90149-0} \end{APACrefDOI}
\PrintBackRefs{\CurrentBib}

\bibitem [\protect \citeauthoryear {%
{Shen}%
\ \protect \BOthers {.}}{%
{Shen}%
\ \protect \BOthers {.}}{%
{\protect \APACyear {2023}}%
}]{%
Shen23:jgr:ELFIN_dropout}
\APACinsertmetastar {%
Shen23:jgr:ELFIN_dropout}%
\begin{APACrefauthors}%
{Shen}, Y.%
, {Artemyev}, A\BPBI V.%
, {Runov}, A.%
, {Angelopoulos}, V.%
, {Liu}, J.%
, {Zhang}, X\BHBI J.%
\BDBL {}{Wilkins}, C.%
\end{APACrefauthors}%
\unskip\
\newblock
\APACrefYearMonthDay{2023}{{\APACmonth{09}}}{}.
\newblock
{\BBOQ}\APACrefatitle {{Energetic Electron Flux Dropouts Measured by ELFIN in the Ionospheric Projection of the Plasma Sheet}} {{Energetic Electron Flux Dropouts Measured by ELFIN in the Ionospheric Projection of the Plasma Sheet}}.{\BBCQ}
\newblock
\APACjournalVolNumPages{Journal of Geophysical Research (Space Physics)}{128}{9}{e2023JA031631}.
\newblock
\begin{APACrefDOI} \doi{10.1029/2023JA031631} \end{APACrefDOI}
\PrintBackRefs{\CurrentBib}

\bibitem [\protect \citeauthoryear {%
Shepherd%
}{%
Shepherd%
}{%
{\protect \APACyear {2014}}%
}]{%
Shepherd2014}
\APACinsertmetastar {%
Shepherd2014}%
\begin{APACrefauthors}%
Shepherd, S\BPBI G.%
\end{APACrefauthors}%
\unskip\
\newblock
\APACrefYearMonthDay{2014}{}{}.
\newblock
{\BBOQ}\APACrefatitle {Altitude-adjusted corrected geomagnetic coordinates: Definition and functional approximations} {Altitude-adjusted corrected geomagnetic coordinates: Definition and functional approximations}.{\BBCQ}
\newblock
\APACjournalVolNumPages{Journal of Geophysical Research: Space Physics}{119}{9}{7501-7521}.
\newblock
\begin{APACrefURL} \url{https://agupubs.onlinelibrary.wiley.com/doi/abs/10.1002/2014JA020264} \end{APACrefURL}
\newblock
\begin{APACrefDOI} \doi{https://doi.org/10.1002/2014JA020264} \end{APACrefDOI}
\PrintBackRefs{\CurrentBib}

\bibitem [\protect \citeauthoryear {%
Shevchenko%
\ \protect \BOthers {.}}{%
Shevchenko%
\ \protect \BOthers {.}}{%
{\protect \APACyear {2010}}%
}]{%
Shevchenko2010}
\APACinsertmetastar {%
Shevchenko2010}%
\begin{APACrefauthors}%
Shevchenko, I\BPBI G.%
, Sergeev, V.%
, Kubyshkina, M.%
, Angelopoulos, V.%
, Glassmeier, K\BPBI H.%
\BCBL {}\ \BBA {} Singer, H\BPBI J.%
\end{APACrefauthors}%
\unskip\
\newblock
\APACrefYearMonthDay{2010}{}{}.
\newblock
{\BBOQ}\APACrefatitle {Estimation of magnetosphere-ionosphere mapping accuracy using isotropy boundary and THEMIS observations} {Estimation of magnetosphere-ionosphere mapping accuracy using isotropy boundary and themis observations}.{\BBCQ}
\newblock
\APACjournalVolNumPages{Journal of Geophysical Research: Space Physics}{115}{A11}{}.
\newblock
\begin{APACrefDOI} \doi{https://doi.org/10.1029/2010JA015354} \end{APACrefDOI}
\PrintBackRefs{\CurrentBib}

\bibitem [\protect \citeauthoryear {%
{Shukhtina}%
, {Dmitrieva}%
\BCBL {}\ \BBA {} {Sergeev}%
}{%
{Shukhtina}%
\ \protect \BOthers {.}}{%
{\protect \APACyear {2014}}%
}]{%
Shukhtina2014}
\APACinsertmetastar {%
Shukhtina2014}%
\begin{APACrefauthors}%
{Shukhtina}, M\BPBI A.%
, {Dmitrieva}, N\BPBI P.%
\BCBL {}\ \BBA {} {Sergeev}, V\BPBI A.%
\end{APACrefauthors}%
\unskip\
\newblock
\APACrefYearMonthDay{2014}{{\APACmonth{02}}}{}.
\newblock
{\BBOQ}\APACrefatitle {{On the conditions preceding sudden magnetotail magnetic flux unloading}} {{On the conditions preceding sudden magnetotail magnetic flux unloading}}.{\BBCQ}
\newblock
\APACjournalVolNumPages{\grl}{41}{4}{1093-1099}.
\newblock
\begin{APACrefDOI} \doi{10.1002/2014GL059290} \end{APACrefDOI}
\PrintBackRefs{\CurrentBib}

\bibitem [\protect \citeauthoryear {%
{Sitnov}%
, {Birn}%
\BCBL {}\ \protect \BOthers {.}}{%
{Sitnov}%
, {Birn}%
\BCBL {}\ \protect \BOthers {.}}{%
{\protect \APACyear {2019}}%
}]{%
Sitnov19}
\APACinsertmetastar {%
Sitnov19}%
\begin{APACrefauthors}%
{Sitnov}, M\BPBI I.%
, {Birn}, J.%
, {Ferdousi}, B.%
, {Gordeev}, E.%
, {Khotyaintsev}, Y.%
, {Merkin}, V.%
\BDBL {}{Zhou}, X.%
\end{APACrefauthors}%
\unskip\
\newblock
\APACrefYearMonthDay{2019}{Jun}{}.
\newblock
{\BBOQ}\APACrefatitle {{Explosive Magnetotail Activity}} {{Explosive Magnetotail Activity}}.{\BBCQ}
\newblock
\APACjournalVolNumPages{\ssr}{215}{4}{31}.
\newblock
\begin{APACrefDOI} \doi{10.1007/s11214-019-0599-5} \end{APACrefDOI}
\PrintBackRefs{\CurrentBib}

\bibitem [\protect \citeauthoryear {%
{Sitnov}%
\ \BBA {} {Schindler}%
}{%
{Sitnov}%
\ \BBA {} {Schindler}%
}{%
{\protect \APACyear {2010}}%
}]{%
Sitnov10}
\APACinsertmetastar {%
Sitnov10}%
\begin{APACrefauthors}%
{Sitnov}, M\BPBI I.%
\BCBT {}\ \BBA {} {Schindler}, K.%
\end{APACrefauthors}%
\unskip\
\newblock
\APACrefYearMonthDay{2010}{{\APACmonth{04}}}{}.
\newblock
{\BBOQ}\APACrefatitle {{Tearing stability of a multiscale magnetotail current sheet}} {{Tearing stability of a multiscale magnetotail current sheet}}.{\BBCQ}
\newblock
\APACjournalVolNumPages{\grl}{37}{}{8102}.
\newblock
\begin{APACrefDOI} \doi{10.1029/2010GL042961} \end{APACrefDOI}
\PrintBackRefs{\CurrentBib}

\bibitem [\protect \citeauthoryear {%
{Sitnov}%
, {Stephens}%
, {Motoba}%
\BCBL {}\ \BBA {} {Swisdak}%
}{%
{Sitnov}%
\ \protect \BOthers {.}}{%
{\protect \APACyear {2021}}%
}]{%
Sitnov21}
\APACinsertmetastar {%
Sitnov21}%
\begin{APACrefauthors}%
{Sitnov}, M\BPBI I.%
, {Stephens}, G.%
, {Motoba}, T.%
\BCBL {}\ \BBA {} {Swisdak}, M.%
\end{APACrefauthors}%
\unskip\
\newblock
\APACrefYearMonthDay{2021}{{\APACmonth{04}}}{}.
\newblock
{\BBOQ}\APACrefatitle {{Data Mining Reconstruction of Magnetotail Reconnection and Implications for Its First-Principle Modeling}} {{Data Mining Reconstruction of Magnetotail Reconnection and Implications for Its First-Principle Modeling}}.{\BBCQ}
\newblock
\APACjournalVolNumPages{Frontiers in Physics}{9}{}{90}.
\newblock
\begin{APACrefDOI} \doi{10.3389/fphy.2021.644884} \end{APACrefDOI}
\PrintBackRefs{\CurrentBib}

\bibitem [\protect \citeauthoryear {%
{Sitnov}%
, {Stephens}%
\BCBL {}\ \protect \BOthers {.}}{%
{Sitnov}%
, {Stephens}%
\BCBL {}\ \protect \BOthers {.}}{%
{\protect \APACyear {2019}}%
}]{%
Sitnov19:jgr}
\APACinsertmetastar {%
Sitnov19:jgr}%
\begin{APACrefauthors}%
{Sitnov}, M\BPBI I.%
, {Stephens}, G\BPBI K.%
, {Tsyganenko}, N\BPBI A.%
, {Miyashita}, Y.%
, {Merkin}, V\BPBI G.%
, {Motoba}, T.%
\BDBL {}{Genestreti}, K\BPBI J.%
\end{APACrefauthors}%
\unskip\
\newblock
\APACrefYearMonthDay{2019}{Nov}{}.
\newblock
{\BBOQ}\APACrefatitle {{Signatures of Nonideal Plasma Evolution During Substorms Obtained by Mining Multimission Magnetometer Data}} {{Signatures of Nonideal Plasma Evolution During Substorms Obtained by Mining Multimission Magnetometer Data}}.{\BBCQ}
\newblock
\APACjournalVolNumPages{Journal of Geophysical Research (Space Physics)}{124}{11}{8427-8456}.
\newblock
\begin{APACrefDOI} \doi{10.1029/2019JA027037} \end{APACrefDOI}
\PrintBackRefs{\CurrentBib}

\bibitem [\protect \citeauthoryear {%
Sitnov%
\ \protect \BOthers {.}}{%
Sitnov%
\ \protect \BOthers {.}}{%
{\protect \APACyear {2017}}%
}]{%
Sitnov2017M}
\APACinsertmetastar {%
Sitnov2017M}%
\begin{APACrefauthors}%
Sitnov, M\BPBI I.%
, Stephens, G\BPBI K.%
, Tsyganenko, N\BPBI A.%
, Ukhorskiy, A\BPBI Y.%
, Wing, S.%
, Korth, H.%
\BCBL {}\ \BBA {} Anderson, B\BPBI J.%
\end{APACrefauthors}%
\unskip\
\newblock
\APACrefYearMonthDay{2017}{}{}.
\newblock
{\BBOQ}\APACrefatitle {Spatial Structure and Asymmetries of Magnetospheric Currents Inferred from High-Resolution Empirical Geomagnetic Field Models} {Spatial structure and asymmetries of magnetospheric currents inferred from high-resolution empirical geomagnetic field models}.{\BBCQ}
\newblock
\BIn{} \APACrefbtitle {Dawn-Dusk Asymmetries in Planetary Plasma Environments} {Dawn-dusk asymmetries in planetary plasma environments}\ (\BPG~199-212).
\newblock
\APACaddressPublisher{}{American Geophysical Union (AGU)}.
\newblock
\begin{APACrefDOI} \doi{10.1002/9781119216346.ch15} \end{APACrefDOI}
\PrintBackRefs{\CurrentBib}

\bibitem [\protect \citeauthoryear {%
{Sitnov}%
, {Tsyganenko}%
, {Ukhorskiy}%
\BCBL {}\ \BBA {} {Brandt}%
}{%
{Sitnov}%
\ \protect \BOthers {.}}{%
{\protect \APACyear {2008}}%
}]{%
Sitnov08}
\APACinsertmetastar {%
Sitnov08}%
\begin{APACrefauthors}%
{Sitnov}, M\BPBI I.%
, {Tsyganenko}, N\BPBI A.%
, {Ukhorskiy}, A\BPBI Y.%
\BCBL {}\ \BBA {} {Brandt}, P\BPBI C.%
\end{APACrefauthors}%
\unskip\
\newblock
\APACrefYearMonthDay{2008}{{\APACmonth{07}}}{}.
\newblock
{\BBOQ}\APACrefatitle {{Dynamical data-based modeling of the storm-time geomagnetic field with enhanced spatial resolution}} {{Dynamical data-based modeling of the storm-time geomagnetic field with enhanced spatial resolution}}.{\BBCQ}
\newblock
\APACjournalVolNumPages{\jgr}{113}{}{7218}.
\newblock
\begin{APACrefDOI} \doi{10.1029/2007JA013003} \end{APACrefDOI}
\PrintBackRefs{\CurrentBib}

\bibitem [\protect \citeauthoryear {%
{Sivadas}%
, {Semeter}%
, {Nishimura}%
\BCBL {}\ \BBA {} {Kero}%
}{%
{Sivadas}%
\ \protect \BOthers {.}}{%
{\protect \APACyear {2017}}%
}]{%
Sivadas17}
\APACinsertmetastar {%
Sivadas17}%
\begin{APACrefauthors}%
{Sivadas}, N.%
, {Semeter}, J.%
, {Nishimura}, Y.%
\BCBL {}\ \BBA {} {Kero}, A.%
\end{APACrefauthors}%
\unskip\
\newblock
\APACrefYearMonthDay{2017}{{\APACmonth{10}}}{}.
\newblock
{\BBOQ}\APACrefatitle {{Simultaneous Measurements of Substorm-Related Electron Energization in the Ionosphere and the Plasma Sheet}} {{Simultaneous Measurements of Substorm-Related Electron Energization in the Ionosphere and the Plasma Sheet}}.{\BBCQ}
\newblock
\APACjournalVolNumPages{Journal of Geophysical Research (Space Physics)}{122}{10}{10,528-10,547}.
\newblock
\begin{APACrefDOI} \doi{10.1002/2017JA023995} \end{APACrefDOI}
\PrintBackRefs{\CurrentBib}

\bibitem [\protect \citeauthoryear {%
{Sorathia}%
\ \protect \BOthers {.}}{%
{Sorathia}%
\ \protect \BOthers {.}}{%
{\protect \APACyear {2020}}%
}]{%
Sorathia20}
\APACinsertmetastar {%
Sorathia20}%
\begin{APACrefauthors}%
{Sorathia}, K\BPBI A.%
, {Merkin}, V\BPBI G.%
, {Panov}, E\BPBI V.%
, {Zhang}, B.%
, {Lyon}, J\BPBI G.%
, {Garretson}, J.%
\BDBL {}{Wiltberger}, M.%
\end{APACrefauthors}%
\unskip\
\newblock
\APACrefYearMonthDay{2020}{{\APACmonth{07}}}{}.
\newblock
{\BBOQ}\APACrefatitle {{Ballooning-Interchange Instability in the Near-Earth Plasma Sheet and Auroral Beads: Global Magnetospheric Modeling at the Limit of the MHD Approximation}} {{Ballooning-Interchange Instability in the Near-Earth Plasma Sheet and Auroral Beads: Global Magnetospheric Modeling at the Limit of the MHD Approximation}}.{\BBCQ}
\newblock
\APACjournalVolNumPages{\grl}{47}{14}{e88227}.
\newblock
\begin{APACrefDOI} \doi{10.1029/2020GL088227} \end{APACrefDOI}
\PrintBackRefs{\CurrentBib}

\bibitem [\protect \citeauthoryear {%
{Stephens}%
\ \BBA {} {Sitnov}%
}{%
{Stephens}%
\ \BBA {} {Sitnov}%
}{%
{\protect \APACyear {2021}}%
}]{%
Stephens&Sitnov21}
\APACinsertmetastar {%
Stephens&Sitnov21}%
\begin{APACrefauthors}%
{Stephens}, G\BPBI K.%
\BCBT {}\ \BBA {} {Sitnov}, M\BPBI I.%
\end{APACrefauthors}%
\unskip\
\newblock
\APACrefYearMonthDay{2021}{{\APACmonth{05}}}{}.
\newblock
{\BBOQ}\APACrefatitle {{Concurrent empirical magnetic reconstruction of storm and substorm spatial scales using data mining and virtual spacecraft}} {{Concurrent empirical magnetic reconstruction of storm and substorm spatial scales using data mining and virtual spacecraft}}.{\BBCQ}
\newblock
\APACjournalVolNumPages{Frontiers in Physics}{9}{}{210}.
\newblock
\begin{APACrefDOI} \doi{10.3389/fphy.2021.653111} \end{APACrefDOI}
\PrintBackRefs{\CurrentBib}

\bibitem [\protect \citeauthoryear {%
{Stephens}%
\ \protect \BOthers {.}}{%
{Stephens}%
\ \protect \BOthers {.}}{%
{\protect \APACyear {2019}}%
}]{%
Stephens19}
\APACinsertmetastar {%
Stephens19}%
\begin{APACrefauthors}%
{Stephens}, G\BPBI K.%
, {Sitnov}, M\BPBI I.%
, {Korth}, H.%
, {Tsyganenko}, N\BPBI A.%
, {Ohtani}, S.%
, {Gkioulidou}, M.%
\BCBL {}\ \BBA {} {Ukhorskiy}, A\BPBI Y.%
\end{APACrefauthors}%
\unskip\
\newblock
\APACrefYearMonthDay{2019}{Feb}{}.
\newblock
{\BBOQ}\APACrefatitle {{Global Empirical Picture of Magnetospheric Substorms Inferred From Multimission Magnetometer Data}} {{Global Empirical Picture of Magnetospheric Substorms Inferred From Multimission Magnetometer Data}}.{\BBCQ}
\newblock
\APACjournalVolNumPages{Journal of Geophysical Research (Space Physics)}{124}{2}{1085-1110}.
\newblock
\begin{APACrefDOI} \doi{10.1029/2018JA025843} \end{APACrefDOI}
\PrintBackRefs{\CurrentBib}

\bibitem [\protect \citeauthoryear {%
{Stephens}%
\ \protect \BOthers {.}}{%
{Stephens}%
\ \protect \BOthers {.}}{%
{\protect \APACyear {2016}}%
}]{%
Stephens16}
\APACinsertmetastar {%
Stephens16}%
\begin{APACrefauthors}%
{Stephens}, G\BPBI K.%
, {Sitnov}, M\BPBI I.%
, {Ukhorskiy}, A\BPBI Y.%
, {Roelof}, E\BPBI C.%
, {Tsyganenko}, N\BPBI A.%
\BCBL {}\ \BBA {} {Le}, G.%
\end{APACrefauthors}%
\unskip\
\newblock
\APACrefYearMonthDay{2016}{{\APACmonth{01}}}{}.
\newblock
{\BBOQ}\APACrefatitle {{Empirical modeling of the storm time innermost magnetosphere using Van Allen Probes and THEMIS data: Eastward and banana currents}} {{Empirical modeling of the storm time innermost magnetosphere using Van Allen Probes and THEMIS data: Eastward and banana currents}}.{\BBCQ}
\newblock
\APACjournalVolNumPages{\jgr}{121}{}{157-170}.
\newblock
\begin{APACrefDOI} \doi{10.1002/2015JA021700} \end{APACrefDOI}
\PrintBackRefs{\CurrentBib}

\bibitem [\protect \citeauthoryear {%
{Stephens}%
\ \protect \BOthers {.}}{%
{Stephens}%
\ \protect \BOthers {.}}{%
{\protect \APACyear {2023}}%
}]{%
Stephens23}
\APACinsertmetastar {%
Stephens23}%
\begin{APACrefauthors}%
{Stephens}, G\BPBI K.%
, {Sitnov}, M\BPBI I.%
, {Weigel}, R\BPBI S.%
, {Turner}, D\BPBI L.%
, {Tsyganenko}, N\BPBI A.%
, {Rogers}, A\BPBI J.%
\BDBL {}{Slavin}, J\BPBI A.%
\end{APACrefauthors}%
\unskip\
\newblock
\APACrefYearMonthDay{2023}{{\APACmonth{02}}}{}.
\newblock
{\BBOQ}\APACrefatitle {{Global Structure of Magnetotail Reconnection Revealed by Mining Space Magnetometer Data}} {{Global Structure of Magnetotail Reconnection Revealed by Mining Space Magnetometer Data}}.{\BBCQ}
\newblock
\APACjournalVolNumPages{Journal of Geophysical Research (Space Physics)}{128}{2}{e2022JA031066}.
\newblock
\begin{APACrefDOI} \doi{10.1029/2022JA031066} \end{APACrefDOI}
\PrintBackRefs{\CurrentBib}

\bibitem [\protect \citeauthoryear {%
{Tsai}%
, {Artemyev}%
, {Zhang}%
\BCBL {}\ \BBA {} {Angelopoulos}%
}{%
{Tsai}%
\ \protect \BOthers {.}}{%
{\protect \APACyear {2022}}%
}]{%
Tsai22}
\APACinsertmetastar {%
Tsai22}%
\begin{APACrefauthors}%
{Tsai}, E.%
, {Artemyev}, A.%
, {Zhang}, X\BHBI J.%
\BCBL {}\ \BBA {} {Angelopoulos}, V.%
\end{APACrefauthors}%
\unskip\
\newblock
\APACrefYearMonthDay{2022}{{\APACmonth{05}}}{}.
\newblock
{\BBOQ}\APACrefatitle {{Relativistic Electron Precipitation Driven by Nonlinear Resonance With Whistler-Mode Waves}} {{Relativistic Electron Precipitation Driven by Nonlinear Resonance With Whistler-Mode Waves}}.{\BBCQ}
\newblock
\APACjournalVolNumPages{Journal of Geophysical Research (Space Physics)}{127}{5}{e30338}.
\newblock
\begin{APACrefDOI} \doi{10.1029/2022JA030338} \end{APACrefDOI}
\PrintBackRefs{\CurrentBib}

\bibitem [\protect \citeauthoryear {%
{Tsyganenko}%
}{%
{Tsyganenko}%
}{%
{\protect \APACyear {1995}}%
}]{%
Tsyganenko95}
\APACinsertmetastar {%
Tsyganenko95}%
\begin{APACrefauthors}%
{Tsyganenko}, N\BPBI A.%
\end{APACrefauthors}%
\unskip\
\newblock
\APACrefYearMonthDay{1995}{{\APACmonth{04}}}{}.
\newblock
{\BBOQ}\APACrefatitle {{Modeling the Earth's magnetospheric magnetic field confined within a realistic magnetopause}} {{Modeling the Earth's magnetospheric magnetic field confined within a realistic magnetopause}}.{\BBCQ}
\newblock
\APACjournalVolNumPages{\jgr}{100}{}{5599-5612}.
\newblock
\begin{APACrefDOI} \doi{10.1029/94JA03193} \end{APACrefDOI}
\PrintBackRefs{\CurrentBib}

\bibitem [\protect \citeauthoryear {%
{Tsyganenko}%
\ \BBA {} {Sitnov}%
}{%
{Tsyganenko}%
\ \BBA {} {Sitnov}%
}{%
{\protect \APACyear {2005}}%
}]{%
Tsyganenko&Sitnov05}
\APACinsertmetastar {%
Tsyganenko&Sitnov05}%
\begin{APACrefauthors}%
{Tsyganenko}, N\BPBI A.%
\BCBT {}\ \BBA {} {Sitnov}, M\BPBI I.%
\end{APACrefauthors}%
\unskip\
\newblock
\APACrefYearMonthDay{2005}{{\APACmonth{03}}}{}.
\newblock
{\BBOQ}\APACrefatitle {{Modeling the dynamics of the inner magnetosphere during strong geomagnetic storms}} {{Modeling the dynamics of the inner magnetosphere during strong geomagnetic storms}}.{\BBCQ}
\newblock
\APACjournalVolNumPages{\jgr}{110}{}{A03208}.
\newblock
\begin{APACrefDOI} \doi{10.1029/2004JA010798} \end{APACrefDOI}
\PrintBackRefs{\CurrentBib}

\bibitem [\protect \citeauthoryear {%
{Tsyganenko}%
\ \BBA {} {Sitnov}%
}{%
{Tsyganenko}%
\ \BBA {} {Sitnov}%
}{%
{\protect \APACyear {2007}}%
}]{%
Tsyganenko&Sitnov07}
\APACinsertmetastar {%
Tsyganenko&Sitnov07}%
\begin{APACrefauthors}%
{Tsyganenko}, N\BPBI A.%
\BCBT {}\ \BBA {} {Sitnov}, M\BPBI I.%
\end{APACrefauthors}%
\unskip\
\newblock
\APACrefYearMonthDay{2007}{{\APACmonth{06}}}{}.
\newblock
{\BBOQ}\APACrefatitle {{Magnetospheric configurations from a high-resolution data-based magnetic field model}} {{Magnetospheric configurations from a high-resolution data-based magnetic field model}}.{\BBCQ}
\newblock
\APACjournalVolNumPages{\jgr}{112}{}{A06225}.
\newblock
\begin{APACrefDOI} \doi{10.1029/2007JA012260} \end{APACrefDOI}
\PrintBackRefs{\CurrentBib}

\bibitem [\protect \citeauthoryear {%
{Wanliss}%
, {Samson}%
\BCBL {}\ \BBA {} {Friedrich}%
}{%
{Wanliss}%
\ \protect \BOthers {.}}{%
{\protect \APACyear {2000}}%
}]{%
Wanliss00}
\APACinsertmetastar {%
Wanliss00}%
\begin{APACrefauthors}%
{Wanliss}, J\BPBI A.%
, {Samson}, J\BPBI C.%
\BCBL {}\ \BBA {} {Friedrich}, E.%
\end{APACrefauthors}%
\unskip\
\newblock
\APACrefYearMonthDay{2000}{{\APACmonth{12}}}{}.
\newblock
{\BBOQ}\APACrefatitle {{On the use of photometer data to map dynamics of the magnetotail current sheet during substorm growth phase}} {{On the use of photometer data to map dynamics of the magnetotail current sheet during substorm growth phase}}.{\BBCQ}
\newblock
\APACjournalVolNumPages{\jgr}{105}{A12}{27673-27684}.
\newblock
\begin{APACrefDOI} \doi{10.1029/2000JA000178} \end{APACrefDOI}
\PrintBackRefs{\CurrentBib}

\bibitem [\protect \citeauthoryear {%
{Wilkins}%
\ \protect \BOthers {.}}{%
{Wilkins}%
\ \protect \BOthers {.}}{%
{\protect \APACyear {2023}}%
}]{%
Wilkins23}
\APACinsertmetastar {%
Wilkins23}%
\begin{APACrefauthors}%
{Wilkins}, C.%
, {Angelopoulos}, V.%
, {Runov}, A.%
, {Artemyev}, A.%
, {Zhang}, X\BPBI J.%
, {Liu}, J.%
\BCBL {}\ \BBA {} {Tsai}, E.%
\end{APACrefauthors}%
\unskip\
\newblock
\APACrefYearMonthDay{2023}{{\APACmonth{10}}}{}.
\newblock
{\BBOQ}\APACrefatitle {{Statistical Characteristics of the Electron Isotropy Boundary}} {{Statistical Characteristics of the Electron Isotropy Boundary}}.{\BBCQ}
\newblock
\APACjournalVolNumPages{Journal of Geophysical Research (Space Physics)}{128}{10}{e2023JA031774}.
\newblock
\begin{APACrefDOI} \doi{10.1029/2023JA031774} \end{APACrefDOI}
\PrintBackRefs{\CurrentBib}

\bibitem [\protect \citeauthoryear {%
Yahnin%
\ \protect \BOthers {.}}{%
Yahnin%
\ \protect \BOthers {.}}{%
{\protect \APACyear {2001}}%
}]{%
Yahnin2001}
\APACinsertmetastar {%
Yahnin2001}%
\begin{APACrefauthors}%
Yahnin, A\BPBI G.%
, Sergeev, V\BPBI A.%
, B\"osinger, T.%
, Sergienko, T\BPBI I.%
, Kornilov, I\BPBI A.%
, Borodkova, N\BPBI L.%
\BDBL {}Skalsky, A\BPBI A.%
\end{APACrefauthors}%
\unskip\
\newblock
\APACrefYearMonthDay{2001}{}{}.
\newblock
{\BBOQ}\APACrefatitle {Correlated Interball/ground-based observations of isolated substorm: The pseudobreakup phase} {Correlated interball/ground-based observations of isolated substorm: The pseudobreakup phase}.{\BBCQ}
\newblock
\APACjournalVolNumPages{Annales Geophysicae}{19}{7}{687--698}.
\newblock
\begin{APACrefDOI} \doi{10.5194/angeo-19-687-2001} \end{APACrefDOI}
\PrintBackRefs{\CurrentBib}

\bibitem [\protect \citeauthoryear {%
{Yahnin}%
, {Sergeev}%
, {Gvozdevsky}%
\BCBL {}\ \BBA {} {Vennerstr{\o}m}%
}{%
{Yahnin}%
\ \protect \BOthers {.}}{%
{\protect \APACyear {1997}}%
}]{%
Yahnin97}
\APACinsertmetastar {%
Yahnin97}%
\begin{APACrefauthors}%
{Yahnin}, A\BPBI G.%
, {Sergeev}, V\BPBI A.%
, {Gvozdevsky}, B\BPBI B.%
\BCBL {}\ \BBA {} {Vennerstr{\o}m}, S.%
\end{APACrefauthors}%
\unskip\
\newblock
\APACrefYearMonthDay{1997}{{\APACmonth{08}}}{}.
\newblock
{\BBOQ}\APACrefatitle {{Magnetospheric source region of discrete auroras inferred from their relationship with isotropy boundaries of energetic particles}} {{Magnetospheric source region of discrete auroras inferred from their relationship with isotropy boundaries of energetic particles}}.{\BBCQ}
\newblock
\APACjournalVolNumPages{Annales Geophysicae}{15}{}{943-958}.
\newblock
\begin{APACrefDOI} \doi{10.1007/s00585-997-0943-z} \end{APACrefDOI}
\PrintBackRefs{\CurrentBib}

\end{thebibliography}

\newpage
\section*{Supporting Information}
\noindent\textbf{Contents of this file}
\begin{enumerate}
\item ELFIN, THEMIS, and MMS flux comparisons
\item SST19 algorithm description  
\item Magnetotail stretching factor and isotropy boundaries for ions and electrons 
\item Figures S1 -- S9
\end{enumerate}

\vspace{0.5cm}
\noindent\textbf{Introduction}

The Supporting Information includes the description of THEMIS and MMS observations and their consistency with ELFIN observations, details of the data mining algorithm SST19, and the SST19 derived estimates for the stretching factor and its equatorial distributions including the corresponding locations for the electron and ion isotropy boundaries. 


\vspace{0.5cm}
\noindent\textbf{ELFIN, THEMIS, and MMS flux comparisons}

Here, we bracket the radial location of ELFIN's projection to the magnetic equator
by comparing its particle fluxes during intervals of isotropic flux measurements to flux data from MMS and THEMIS spacecraft. We select ELFIN spectra with flux at 50--100keV closest to what THEMIS and MMS observed. We use omni-directional flux averages from THEMIS and MMS (see references with spacecraft instruments in the main text). Figures \ref{figS1}(a--d) depict magnetic field and electron flux observations obtained from THEMIS-E and MMS during 05:00--11:00 UT. In this interval, THEMIS-E was located close to the equator, moving tailward from $8R_E$ to $11R_E$. MMS was located at a radial distance of about $16R_E$ well below the equator ($|B_x|>B_z$) and mostly near the plasma sheet boundary layer which maps to equatorial locations much farther tailward than $16R_E$. Figure \ref{figS1}(g) compares electron flux measurements from THEMIS at $8R_E$ and $10R_E$ (black lines), MMS at $16R_E$ (red lines), and ELFIN (at the ionosphere) both near the IBe (blue lines) and far poleward of it (gray line). The comparison between ELFIN and THEMIS fluxes suggests that the IBe source population is outside $8R_E$ and inside of, or near $10R_E$. The electron flux measured by ELFIN well poleward of the IBe (gray line) at 08:45 UT is comparable to the MMS flux measured at $\sim 16R_E$ and mapping to the plasma sheet equator near (though still tailward of) $\sim 16R_E$. In summary, the above ELFIN/MMS/THEMIS comparisons demonstrate that the IBe observed by ELFIN maps to $\sim10R_E$ and that the ELFIN plasma sheet measurements project to distances tailward of $16R_E$.

Figure~\ref{figS1}(e) shows the ion flux measured by MMS. For the two intervals, around $\sim$06:00 and 10:00 UT, MMS was in the local plasma sheet and was closest to the local equatorial plane to measure ion fluxes above the noise level. The MMS ion spectra for these two times are shown in Fig. \ref{figS1}(f) for comparison with ELFIN's. As expected from electron spectra comparisons, MMS ion fluxes from equatorial distances at or beyond $16R_E$ are well below ELFIN ion fluxes within the electron isotropy boundary, confirming that the ELFIN IBe was mapped to the equator well earthward of $\sim 16R_E$ radial distances. Figure \ref{figS1}(f,g) determines the equatorial projection of the IBe (and IBi, that is equatorward of, or maps earthward of the IBe) in the near-Earth plasma sheet at or around $8R_E$ to $10R_E$, the likely transition region between the plasma sheet and the outer radiation belt at that time.

\vspace{0.5cm}
\noindent\textbf{SST19 algorithm description}

In this study, we employ the SST19 empirical magnetic field reconstruction algorithm (\textit{Stephens et al.,} 2019, 2023; \textit{Stephens and Sitnov, 2021}) to determine the location of IBs within the magnetotail during a substorm. In particular, the version of SST19 used here follows that of \textit{Stephens et al.,} (2023), with three modifications detailed below. First, 29 additional months of MMS data have been added to the space magnetometer archive. Secondly, a new formulation of the spatially varying thickness for the magnetotail current sheet is utilized. Third, to better resolve the magnetic field in the inner magnetosphere and the transition region, the ``merged resolution" procedure (\textit{Stephens and Sitnov, 2021}) is employed.

The SST19 algorithm consists of a data mining (DM) and a fitting component. The DM part characterizes the storm/substorm state of the magnetosphere using a 5-D set of global parameters $\mathbf{G}(t)=(G_1-G_5)$: composed of the solar wind electric field, $vB_s^\mathrm{IMF}$, (where $v$ is the solar wind speed and $B_s=-B_z^\mathrm{IMF}$ when the northward component of the IMF is negative, $B_z^\mathrm{IMF}<0$, and $B_s=0$ otherwise), the SuperMAG pressure-corrected storm, $\textit{SMRc}$, and substorm, $\textit{SML}$, indices, as well as their time derivatives. These parameters are smoothed in time over substorm and storm scales, standardized by dividing by their standard deviations, and sampled at a 5-min cadence as is detailed in \textit{Stephens et al.,} (2019, 2023).

At every moment of interest, $t=t^{(q)}$, the historical archive of space magnetometer observations is mined in the 5D state-space to select other moments, termed nearest-neighbors (NNs), whose global parameters, $\mathbf{G}$, are closest to the query point $\mathbf{G}^{(q)}$.  The archive of magnetometer data ($\sim 9.2\cdot 10^6$ records averaged to 5- and 15-min cadences spanning the years 1995--2023) employed here is similar to that \textit{Stephens et al.,} (2023) with the only difference being that the MMS portion was extended to include data through the end of May 2023, thereby adding 29 months of MMS data. The chosen number of NNs, $k_\mathrm{NN}$, must be small enough to ensure sensitivity to the specific event being reconstructed while large enough to avoid overfitting. Here, as with prior SST19 studies, $k_\mathrm{NN}= 32,000$ which corresponds to $S_\mathrm{NN}\sim 9\cdot 10^4$ magnetometer records or approximately $1\%$ of the entire archive, which can include the few records available at $t=t^{(q)}$. $S_\mathrm{NN}$ is not constant and is larger than $k_\mathrm{NN}$ as there tends to be more than one magnetometer record for any given NN. The NN farthest from $\mathbf{G}^{(q)}$ defines the radius, $R_\mathrm{NN}$, of the NN hypersphere, and distances are computed using the Euclidean distance metric such that $|\mathbf{G}-\mathbf{G}^{(q)}| \leq R_\mathrm{NN}$. $R_\mathrm{NN}$ likewise is not constant and generally increases during times with greater activity.

This instance-based set of $S_\mathrm{NN}$ magnetometer records is then used to fit the analytical description of the magnetic field model represented by the sum: $\mathbf{B}=\mathbf{B}_\mathrm{int}+\mathbf{B}_\mathrm{eq}+\mathbf{B}_\mathrm{FAC}+\mathbf{B}_\mathrm{MP}$, where the internal field, $\mathbf{B}_\mathrm{int}$, is not part of the reconstruction but prescribed using the IGRF model (\textit{Alken et al., } 2021). The reconstructed fields $\mathbf{B}_\mathrm{eq}$, $\mathbf{B}_\mathrm{FAC}$, and $\mathbf{B}_\mathrm{MP}$ are generated by the equatorial, field-aligned (FAC), and magnetopause currents, respectively. Since the resulting reconstruction is specific to the query-time, $t^{(q)}$, it is not universal and is thereby ``event-oriented". This learning process is called instance-based, in contrast to conventional empirical reconstructions (e.g., \textit{Tsyganenko and Sitnov}, 2005, and refs. therein) which are called ``model-based" using machine learning terminology. The fitting element finds the optimal values for the free parameters defining the model's analytical structure by minimizing the root-mean-square difference between the modeled magnetic field and the set of $S_{NN}$ magnetometer records. The result is the reconstructed magnetic field for time $t^{(q)}$: $\mathbf{B}(\mathbf{r}, t=t^{(q)})$. The DM and fitting components are repeated for each time step, at a 5-min cadence, to resolve the dynamical evolution of the magnetospheric magnetic field.

A fundamental advantage of the SST19 approach compared to conventional empirical reconstructions is the description of the magnetospheric currents using basis function expansions for the corresponding magnetic fields rather than custom-made modules with variable amplitudes. In particular, the magnetic field of the equatorial current system, $\mathbf{B}_\mathrm{eq}$, is based on the general solution for Amp\`{e}re's equation for a thin current sheet (CS) in cylindrical coordinates $(\rho, \phi, z)$ taking the form (\textit{Tsyganenko and Sitnov}, 2007):

\begin{equation}
\mathbf{B}_{\mathrm{sheet}}(\rho,\phi,z)=
\displaystyle\sum_{n=1}^{N} a_{0n}^\mathrm{(s)} \mathbf{B}_{0n}^\mathrm{(s)} +
\displaystyle\sum_{m=1}^{M} \displaystyle\sum_{n=1}^{N}
(a_{mn}^\mathrm{(o)}\mathbf{B}_{mn}^\mathrm{(o)}
+
 a_{mn}^\mathrm{(e)}\mathbf{B}_{mn}^\mathrm{(e)}),
\label{sheet eq}
\end{equation}
where $\mathbf{B}_{\alpha\beta}^{(\gamma)}$ are basis functions with axial, odd (sine), and even (cosine) symmetry, while $a_{\alpha\beta}^{(\gamma)}$ are the amplitude coefficients. This solution possesses a characteristic CS half-thickness represented by the variable $D$.
An example of the basis functions used in eq.~(\ref{sheet eq}) can be given by the azimuthal component, $A_{\phi}$, of the vector potential corresponding to the first group of basis functions: $\mathbf{B}_{0n}^{(s)}$: $(A_{\phi})_{0n}^\mathrm{(s)}=J_1(k_n\rho)\exp{(-k_n\sqrt{z^2+D^2})}$, where $J_1$ is the Bessel function of the first order, $k_n=n/R_0$, and $R_0$ is the radial scale set to $20R_E$, corresponding to the largest mode in the radial expansion. The variables $R_0$, $N$, and $M$ are fixed because they determine the adopted spatial resolution of the equatorial current described by eq.~(\ref{sheet eq}). Other variables, such as the coefficients $a_{\alpha\beta}^{(\gamma)}$ and the CS thickness $D$, are determined by fitting the model to data. Thus, the spatial resolution of such an expansion is determined by the number of terms in eq.~(\ref{sheet eq}) and can be increased to any desired level, commensurate with the data availability. Each element in eq.~(\ref{sheet eq}) is independently shielded (has its subsystem of Chapman-Ferraro-type currents at the magnetopause contributing to $\mathbf{B}_\mathrm{MP}$). 

To account for the dependence on the solar wind dynamic pressure, $P_\mathrm{dyn}$, the scaling coefficients $a_{\alpha\beta}^{(\gamma)}$ are split into two groups $a_{\alpha\beta}^{(\gamma)} \to a_{\alpha\beta}^{(\gamma)}+a_{\alpha\beta}^{(\gamma)}\sqrt{P_\mathrm{dyn}}$, doubling the number of coefficients in eq.~(\ref{sheet eq}). To account for seasonal and diurnal variations of the Earth's dipole tilt angle $\Psi$, resulting in a periodic transverse motion and large-scale deformation of the tail current sheet as well as the CS warping and twisting effects, the equatorial magnetic field is deformed (e.g., \textit{Tsyganenko and Sitnov}, 2005, and refs. therein). These deformations introduce three free nonlinear parameters: the hinging distance $R_\mathrm{H}$, the warping parameter $G$, and the twisting parameter $T_\mathrm{W}$, which are determined during the fit.

SST19 employs two independent equatorial current sheets described by eq.~(\ref{sheet eq}) with the same structure but different CS half-thickness parameters $D$ and $D_\mathrm{TCS}$ separated by their spatial sizes to take into account the buildup and decay of ion-scale thin current sheets (TCS) during substorms (\textit{Sergeev et al.,} 2011). 
Moreover, while the thicker CS adopts a spatially constant half-thickness defined using a single free nonlinear parameter $D=\mathrm{const}$, the structure of the TCS utilizes a spatially varying half-thickness, $D_\mathrm{TCS}(\rho)$, based on the distance from Earth, $\rho=\sqrt{x^2+y^2}$, following similar approaches in \textit{Tsyganenko and Sitnov}, (2007) and \textit{Stephens et al.,} (2023). Here, we employ a new flexible TCS thickness model:

\begin{equation}
D_\mathrm{TCS}(\rho)=d_\mathrm{c}\tanh{[[d_\mathrm{thr}+D_0\tanh(\beta\rho)+\alpha\exp{(-\varepsilon \rho/2)}\cos{(3\varepsilon \rho/2)}]/d_c]},
\label{eq2}
\end{equation}
introducing four free nonlinear parameters $D_0$, $\alpha$, $\beta$, and $\varepsilon$, whereas $d_\mathrm{thr}=0.2R_E$ and $d_\mathrm{c}=3.0R_E$ are fixed to constrain the TCS thickness within a physically valid range.

The last element employed for the description of the magnetic field from the equatorial current system is the ``merged resolution" version of the SST19
(\textit{Stephens and Sitnov}, 2021). The ``merged resolution" version addresses a shortcoming of the SST19 approach: resolving different regions of the equatorial current systems requires different spatial resolutions for the same set of NNs, that is, there is not a single value for $(M,N)$ that concurrently resolves all the equatorial currents. This is caused by the disparate density of the available spacecraft magnetometer data and the inherently different spatial scales of the equatorial current systems. For example, adequate resolution of the inner magnetosphere and the transition region ($r < 12 R_E$) requires around 20 radial expansions ($N=20$ in eq.~(1)) (\textit{Stephens et al.,} 2016). However, applying this resolution to the near-tail region ($r > 12 R_E$), where the data density is sparser, results in signatures of overfitting. This results in using one spatial resolution for studies of the inner magnetosphere and the transition region, $(M,N)=(3,20)$ or $560$ equatorial expansions in eq.~(1), and another for resolving the near-tail region, $(M,N)=(6,8)$ or $416$ equatorial expansions. The ``merged resolution" version rectifies this issue by reconstructing the inner magnetosphere and the transition region using the higher spatial resolution ($(M,N)=(3,20)$), reconstructing the magnetotail using the lower resolution ($(M,N)=(6,8)$), and then merging these two results into one coherent picture that concurrently resolves both regions. The merging procedure fits a composite resolution architecture, $(M,N)=(6,20)$, to magnetic field records constructed by randomly sampling the two other magnetic field reconstructions within their respective spatial domains, as is described in detail in \textit{Stephens and Sitnov}, (2021).

The magnetic field of the FAC system, $\mathbf{B}_\mathrm{FAC}$, is presented in the form of multiple elementary current blocks similar to eq.~(\ref{sheet eq}) and described in more detail in \textit{Sitnov et al.}, (2017). The number of blocks is $N_\mathrm{FAC}=16$, which introduces 16 more linear amplitude coefficients. Two additional free nonlinear parameters, $\kappa_\mathrm{R1}$ and $\kappa_\mathrm{R2}$, spatially rescale the higher latitude and lower latitude systems respectively, enabling the FAC ovals to expand and contract with changing activity levels. As with the equatorial system, each FAC system is shielded by adding additional terms to $\mathbf{B}_\mathrm{MP}$, as described in previous works. Other key variables of the SST19 algorithm include the NN weighting index, $\sigma=0.3$, used to assign different weights to the NNs depending on their distance to the query point within the $R_\mathrm{NN}$ hypersphere (see \textit{Stephens et al.} (2023) for more detail), and the maximum radial distance of magnetometer records included in the reconstructions, $r_\mathrm{max}=36 R_E$, which are the same values used in \textit{Stephens et al.} (2023).

This analytical structure is fit by minimizing the weighted root-mean-square difference between the SST19 magnetic field and the magnetic field of the set of $S_\mathrm{NN}$ magnetometer records. Specifically, the linear amplitude coefficients are found by applying the standard singular value decomposition method for linear regression, while the 10 free nonlinear parameters ($D$, $D_0$, $\alpha$, $\beta$, $\varepsilon$, $R_\mathrm{H}$, $G$, $T_\mathrm{W}$, $\kappa_\mathrm{R1}$, $\kappa_\mathrm{R2}$ ) are solved using the Nelder-Mead downhill simplex method (\textit{Press et al.}, 1992). The linear solver is nested within the nonlinear one such that a new set of amplitude coefficients is found whenever the values of a nonlinear parameter are adjusted. This process repeats for 80 iterations.

In summary, the configuration of the SST19 model employed here follows that of \textit{Stephens et al.,} (2023), which itself is a modest revision to the first \textit{Stephens et al.,} (2019) version, with three changes: (1) an additional 29 months of MMS magnetometer records included in the space magnetometer archive, (2) the new formulation for the spatially varying TCS structure described by eq.~(2), and (3) the employment of the ``merged resolution" description of the equatorial current system detailed in \textit{Stephens and Sitnov} (2021).

\noindent\textbf{Magnetotail stretching factor and isotropy boundaries for ions and electrons}

To assess the stretching of the tail magnetic field, which may cause chaos in particle orbits and isotropize their distributions as detected by ELFIN at low-altitudes, we calculate the parameter $G^*=B_z^2/(\mu_0j)$, here termed the magnetotail stretching factor. Lower values of $G^*$ indicate a more stretched magnetotail. The value of the stretching factor for an ideal infinitely thin current sheet, with antiparallel magnetic field lines above and below the sheet, is zero. In contrast, the magnetic equator of a purely dipolar field, where the current density is zero, has an infinitely large stretching factor. Its equatorial distributions (along with the corresponding distributions of meridional and equatorial currents as well as the equatorial $B_z$ field) for several moments in the 19 August 19 2022 substorm are shown in Figures~\ref{figS2} and \ref{figS3}. The isotropization is expected to appear (e.g., \textit{Sergeev et al.}, 2018, and refs. therein) when $G^*<8G$, where $G_{e,p}=m_{e,p}V_{e,p}/e$ is the particle rigidity, $m_{e,p}$ are masses of electrons and protons, and $V_{e,p}$ their velocities. The specific values used in the plots correspond to 520 keV electrons and protons. In constructing the equatorial distributions in Figures~2, 3, \ref{figS2}, and~\ref{figS3}, the dipole tilt and twisting deformations are ignored by setting $\Psi=T_\mathrm{W}=0$.

Figures~\ref{figS4} and \ref{figS5} show the equatorial $B_z$ and $G^*$ mapped to ELFIN's altitude on an $\textit{MLT}$-$\textit{MLAT}$ grid. These plots facilitate the subsequent comparison of ELFIN-observed IBs with those inferred from the SST19 magnetic field reconstruction. Also, to provide the comparison of these reconstructions with similar results from other missions (e.g., \textit{Sergeev et al}., 2018), the analogs of Figures~\ref{figS4} and \ref{figS5} on an $\textit{MLT}$-$\textit{AACGMLAT}$ grid (\textit{Shepherd} 2014) are provided in Figures~\ref{figS6} and \ref{figS7}.  Note, that the higher latitude portions of the ELFIN orbits often map to either open field lines or field lines that intersect the magnetic equator beyond the $r = 20 R_E$. Any IBs inferred using $G^*$ values that map to $r > 20 R_E$ are not considered in the main text. In addition, Figure~\ref{figS8} presents validation of SST19 reconstructions using the magnetic field from THEMIS and MMS. 

Finally, to make sure that our reconstruction algorithm is reliable, we provided in Figure~\ref{figS9} (in the format similar to Figure~2) the reconstruction of three more IB events, one of which is another V-like pattern event shown in Figure~8 of \textit{Artemyev et al.} (2023) (2022-08-11). Two other events describe the IBe detected by ELFIN on June 22, 2021, when the ion data was not available, and an event with very strong activity (2022-09-05) when IBi and IBe came particulalrly close to each other because of the very strong tail current and its strong radial gradient.

\renewcommand{\thefigure}{S\arabic{figure}}

\setcounter{figure}{0}

\vspace{0.5cm}
\noindent\textbf{}

\begin{figure}
\centering
 \noindent\includegraphics[width=0.6\textwidth]{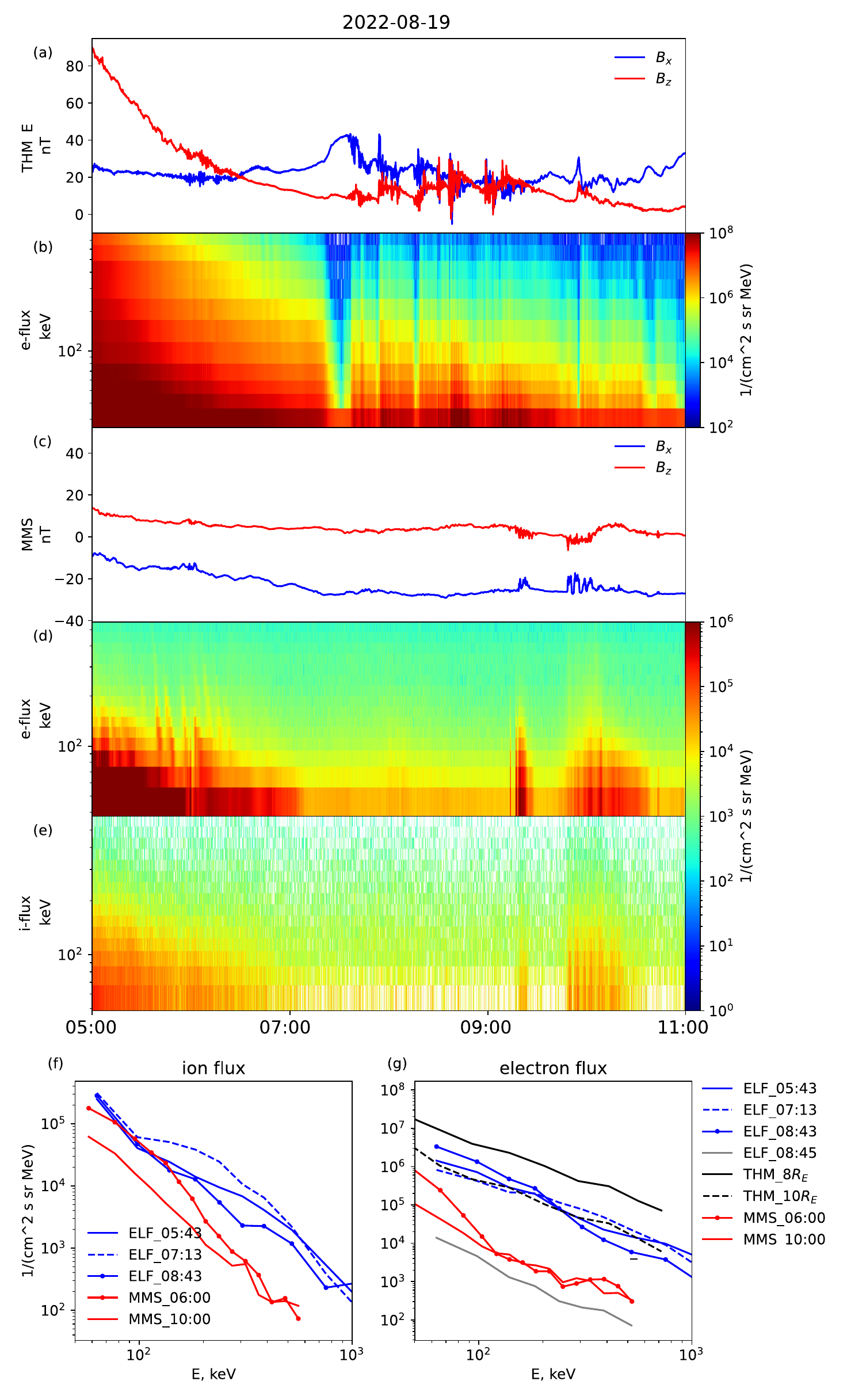}
\caption{Observations from THEMIS and MMS: (a,c) Magnetic field measurements ($B_x$ and $B_z$) in GSM coordinates. (b,d) Electron fluxes between $50$~keV to $600$~keV. (e) Ion fluxes from $50$~keV to $600$~keV. (f) Ion spectra measured by ELFIN and MMS at the times indicated. (g) Electron spectra measured by ELFIN, THEMIS, and MMS at the times indicated. Time moments of ELFIN, THEMIS, and MMS spectra are shown in the panels; these moments indicate the center time for $\pm 1.5$s averaging ELFIN data and $\pm15$min averaging MMS and THEMIS data.
\label{figS1}}
\end{figure}

\begin{figure}
 \centering
 \noindent\includegraphics[width=34pc]{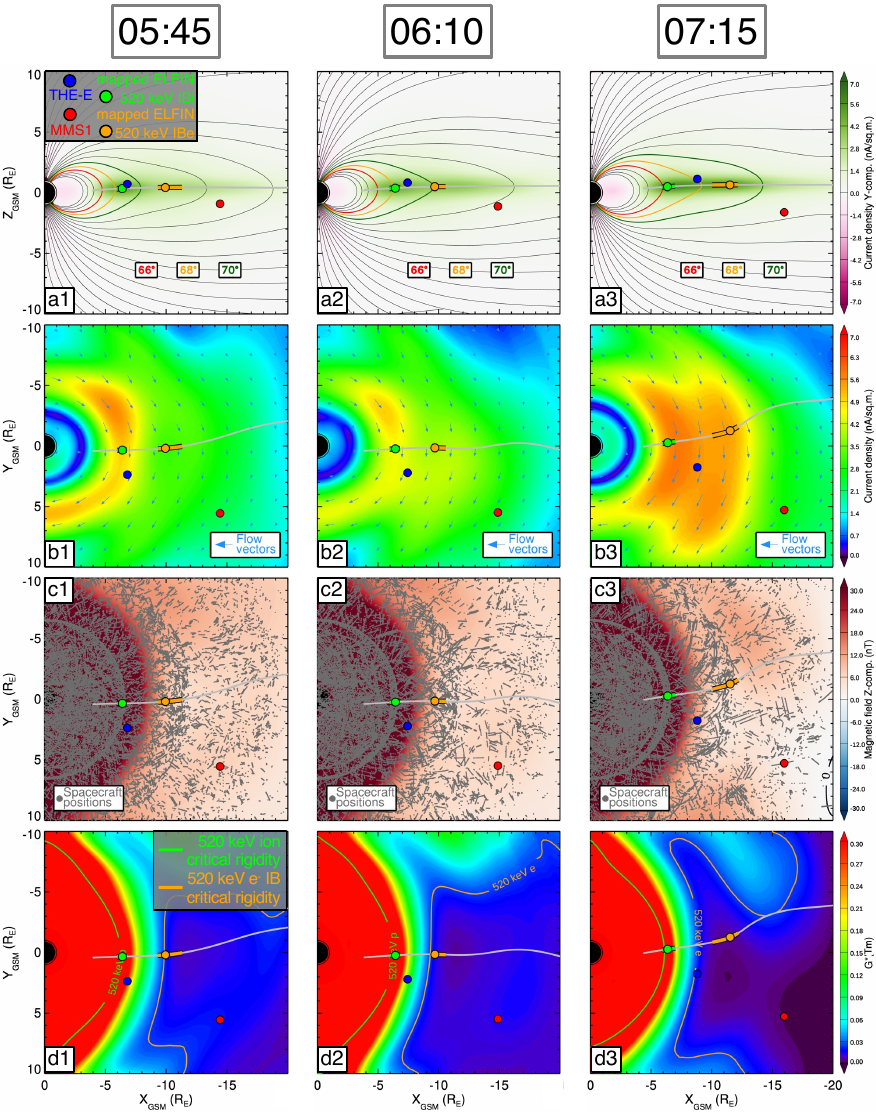}
  \caption{(Caption next page.)}
 \label{figS2}
\end{figure}
\addtocounter{figure}{-1}
\begin{figure} [t!]
 \caption{DM-based empirical reconstructions of the magnetotail when ELFIN observed IBs during the 19~August 2022 substorm. The times approximately correspond to the first three lines from Figure~1 but are rounded to the nearest 5-min based on the SST19's time cadence. (a1--a3) Meridional slices ($y=0$) of the color-coded y-component of the electric current density, $j_y$, with green (pink) corresponding to current flowing out of (into) the page. Magnetic field lines (black), seeded at every $2~^{\circ}$ $\textit{MLAT}$, are overplotted with selected field lines at $\textit{MLAT}=66^{\circ}, 68^{\circ}, 70^{\circ}$ highlighted. (b1--b3) Color-coded equatorial distributions of the electric current density, $j$, with arrows overplotted to indicate the direction of vector current density, $\mathbf{j}$. (c1--c3) Color-coded equatorial distributions of the z-component of the magnetic field, $B_z$, with grey dots overplotted to indicate the locations, projected to the $x$-$y$ plane, of the spacecraft magnetometer observations identified using the $K_\mathrm{NN}$ procedure and used to fit the analytical description of the magnetic field. (d1--d3) Color-coded equatorial distributions of the stretching factor $G^*=B_z^2/(\mu_0 j)$. The critical values of this parameter corresponding to the IB for 520~keV electrons and protons are shown by the orange and green contours respectively. The projections of the locations of the THEMIS~E and MMS1 at the indicated times are overplotted by the colored circles in each panel. The location of ELFIN mapped to the magnetic equator at the time when it observed the 520~keV IBi and IBe are overplotted in green and orange circles respectively.}
\end{figure}

\begin{figure}
 \centering
 \noindent\includegraphics[width=34pc]{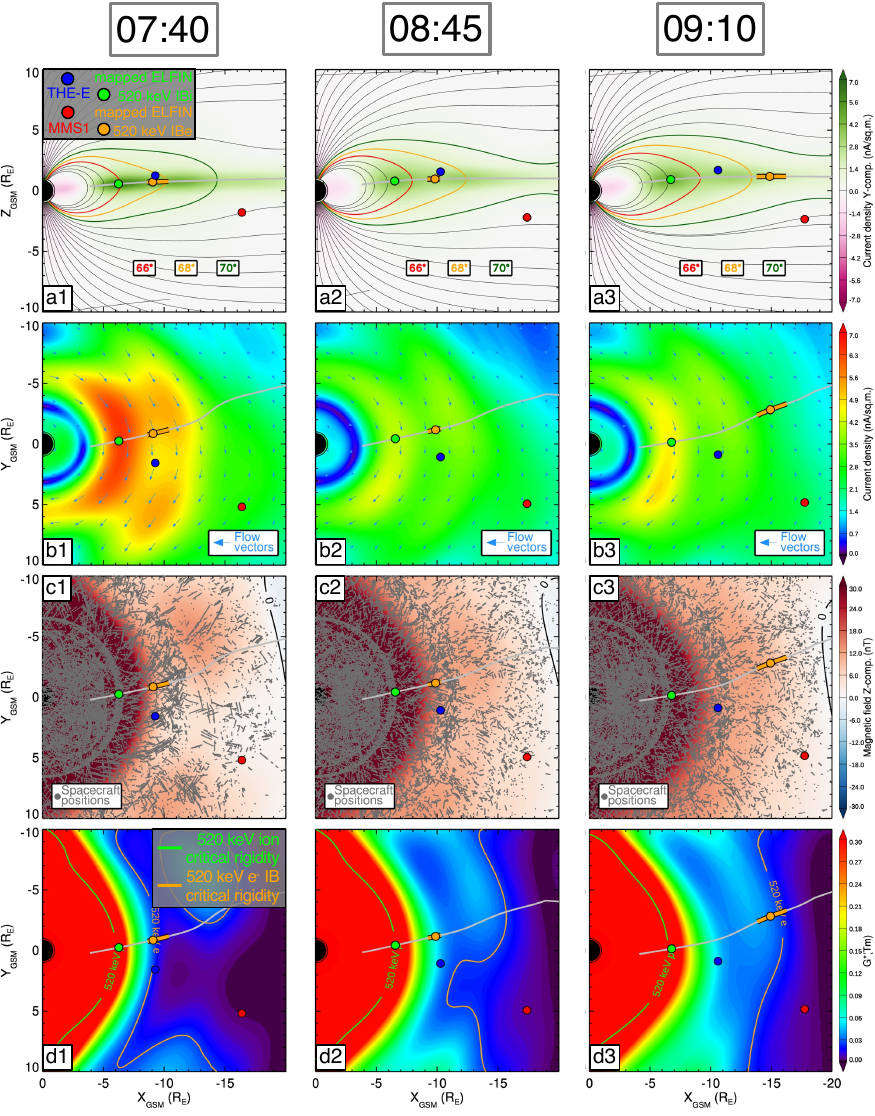}
 \caption{DM-based empirical reconstructions of the magnetotail when ELFIN observed IBs during the 19~August 2022 substorm. The times approximately correspond to the last three lines from Figure~1 but are rounded to the nearest 5-min based on the SST19's time-cadence. The panels are the same as Figure~\ref{figS1} but are at different times.}
  \label{figS3}
\end{figure}

\begin{figure}
 \centering
 \noindent\includegraphics[width=0.9\textwidth]{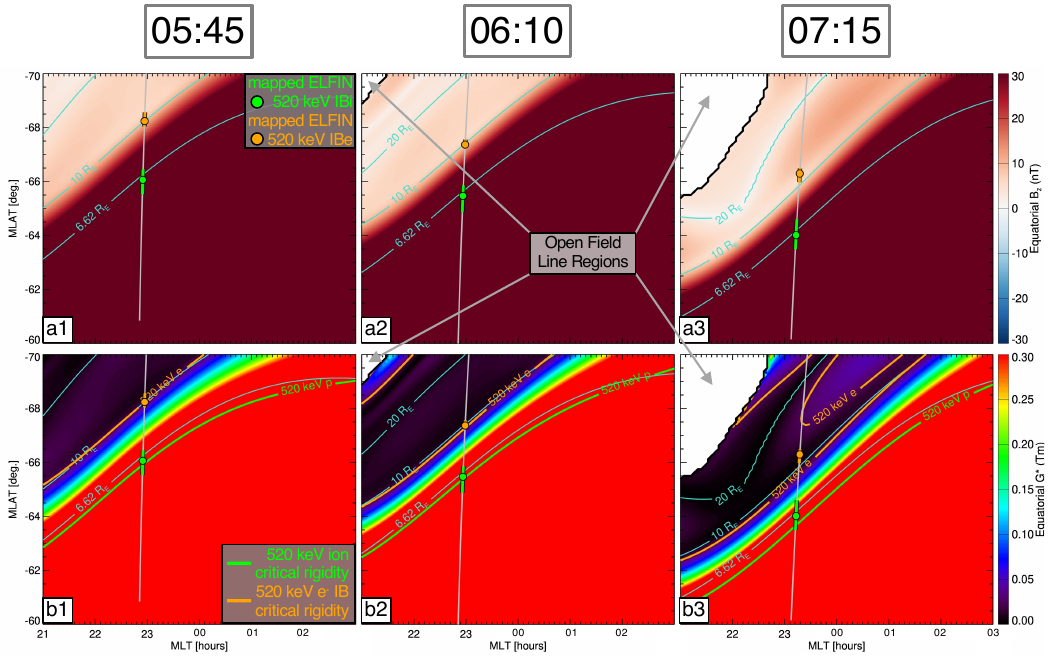}
 \caption{DM-based empirical reconstructions of the magnetotail mapped to the altitude of ELFIN's orbit during the 19~August 2022 substorm. The times approximately correspond to the first three lines from Figure~1 but are rounded to the nearest 5-min based on the SST19's time-cadence. (a1--a3) The color-coded equatorial distribution of the z-component of the magnetic field mapped to $328$ km altitude in the midnight sector. (b1--b3) The color-coded equatorial distributions of the stretching factor $G^*=B_z^2/(\mu_0 j)$ mapped to $328$ km altitude in the midnight sector. ELFIN's orbit is indicated by the grey line and its locations where it observed the 520~keV IBi and IBe are shown by the green and orange circles respectively. The turquoise contours indicate the radial distance of the magnetic equator position to which the field lines map.}
  \label{figS4}
\end{figure}

\begin{figure}
 \centering
 \noindent\includegraphics[width=0.9\textwidth]{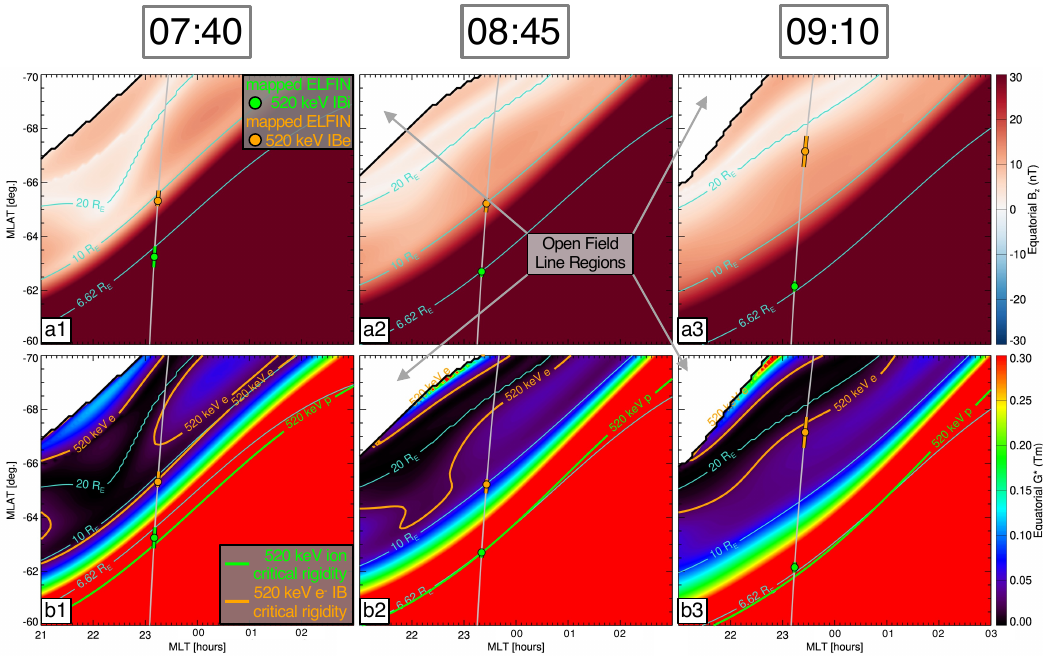}
 \caption{DM-based empirical reconstructions of the magnetotail mapped to the altitude of ELFIN's orbit during the 19~August 2022 substorm. The times approximately correspond to the last three lines from Figure~1 but are rounded to the nearest 5-min based on the SST19's time-cadence. The panels are the same as Figure~\ref{figS3} but are at different times.}
  \label{figS5}
\end{figure}

\begin{figure}
 \centering
 \noindent\includegraphics[width=0.9\textwidth]{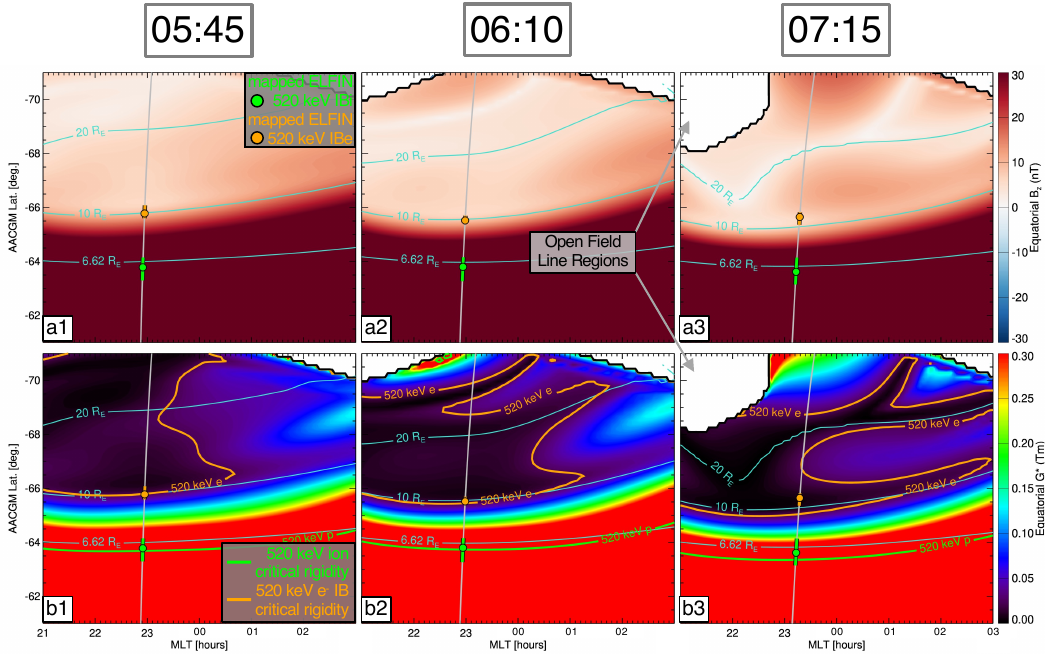}
 \caption{DM-based empirical reconstructions of the magnetotail mapped to the altitude of ELFIN's orbit during the 19~August 2022 substorm. The panels are the same as Figure~S3 but the $y$-axis now uses AACGM latitudes.}
  \label{figS6}
\end{figure}

\begin{figure}
 \centering
 \noindent\includegraphics[width=0.9\textwidth]{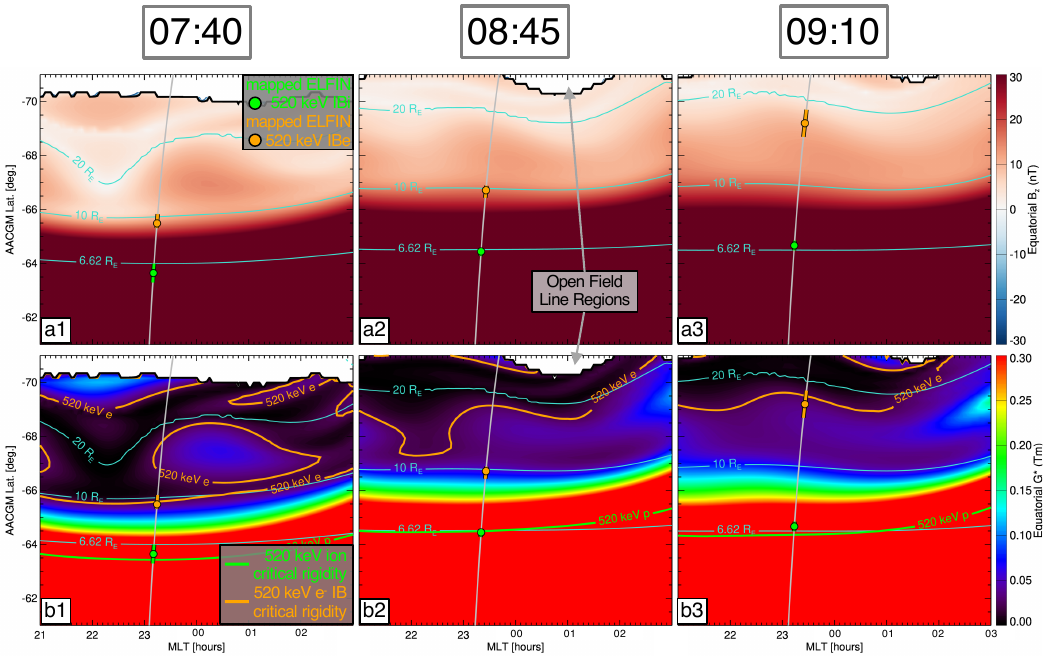}
 \caption{DM-based empirical reconstructions of the magnetotail mapped to the altitude of ELFIN's orbit during the 19~August 2022 substorm. The panels are the same as Figure~S4 but the $y$-axis now uses AACGM latitudes.}
  \label{figS7}
\end{figure}

\begin{figure}
 \centering
 \noindent\includegraphics[width=40pc]{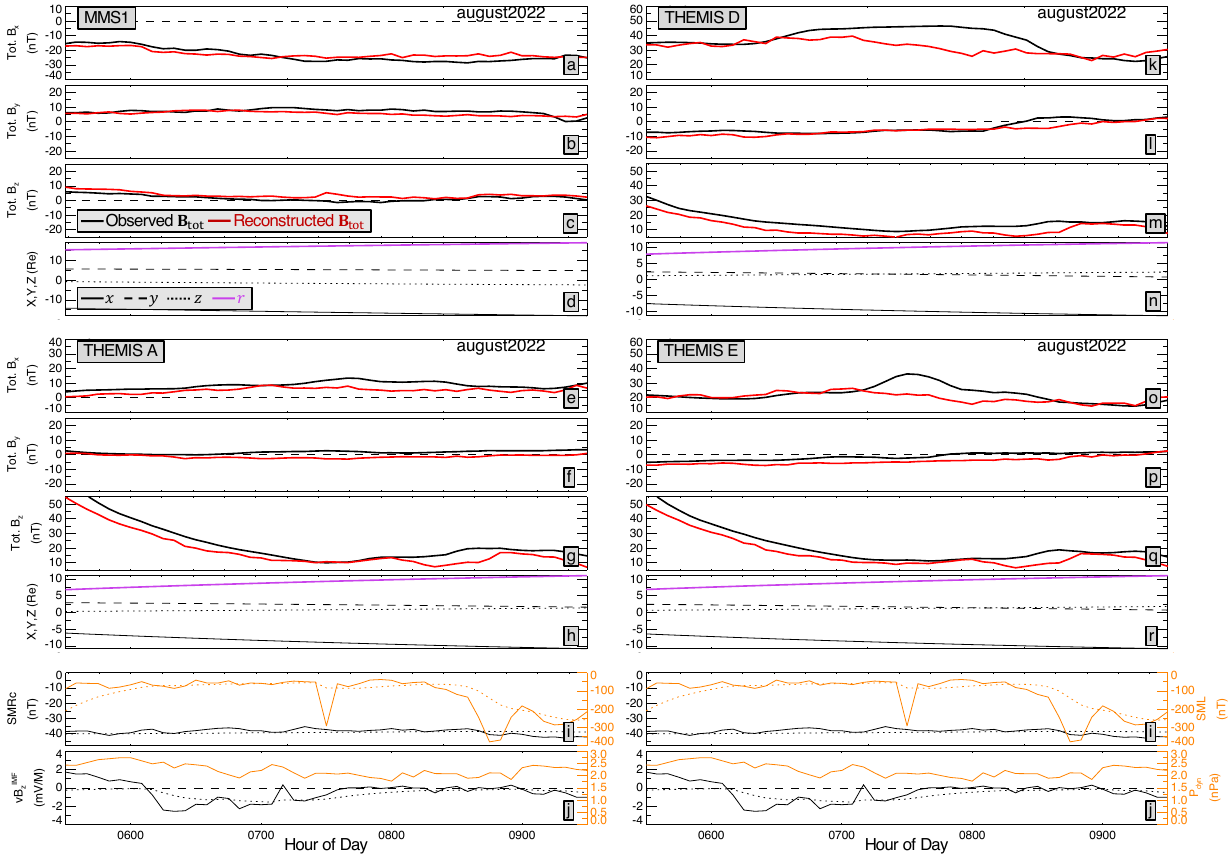}
 \caption{Validation of the DM-based empirical reconstructions of the 19~August 2022 substorm. (a--c) Component-wise comparison between the observed total magnetic field, $\mathbf{B}_\mathrm{tot}$, (black line) averaged to a 5-min cadence and its modeled value (red line) also at a 5-min cadence for the MMS1 spacecraft in GSM coordinates. (d) The MMS1 ephemeris showing the $x$ (solid line), $y$ (dashed line), $z$ (dotted line), and radial distance $r$ (purple line) in GSM coordinates. (e--h) The same as panels (a--d) except for the THEMIS~A spacecraft. (i) The SuperMAG pressure-corrected storm index $\textit{SMRc}$ (black line) and substorm index $\textit{SML}$ (orange line). Their smoothed values, used in the $K_\mathrm{NN}$ procedure, are indicated by the dashed lines. (j) The solar wind electric field parameter $vB_z^\mathrm{IMF}$ (black line) and dynamic pressure $P_\mathrm{dyn}$ (orange line). The smoothed value of $vB_s^\mathrm{IMF}$, used in the $K_\mathrm{NN}$ procedure, is indicated by the dashed line. (k--n) The same as panels (a--d) except for the THEMIS~D spacecraft. (o--r) The same as panels (a--d) except for the THEMIS~E spacecraft.}
  \label{figS8}
\end{figure}

\begin{figure}
 \centering
\noindent\includegraphics[width=38pc]{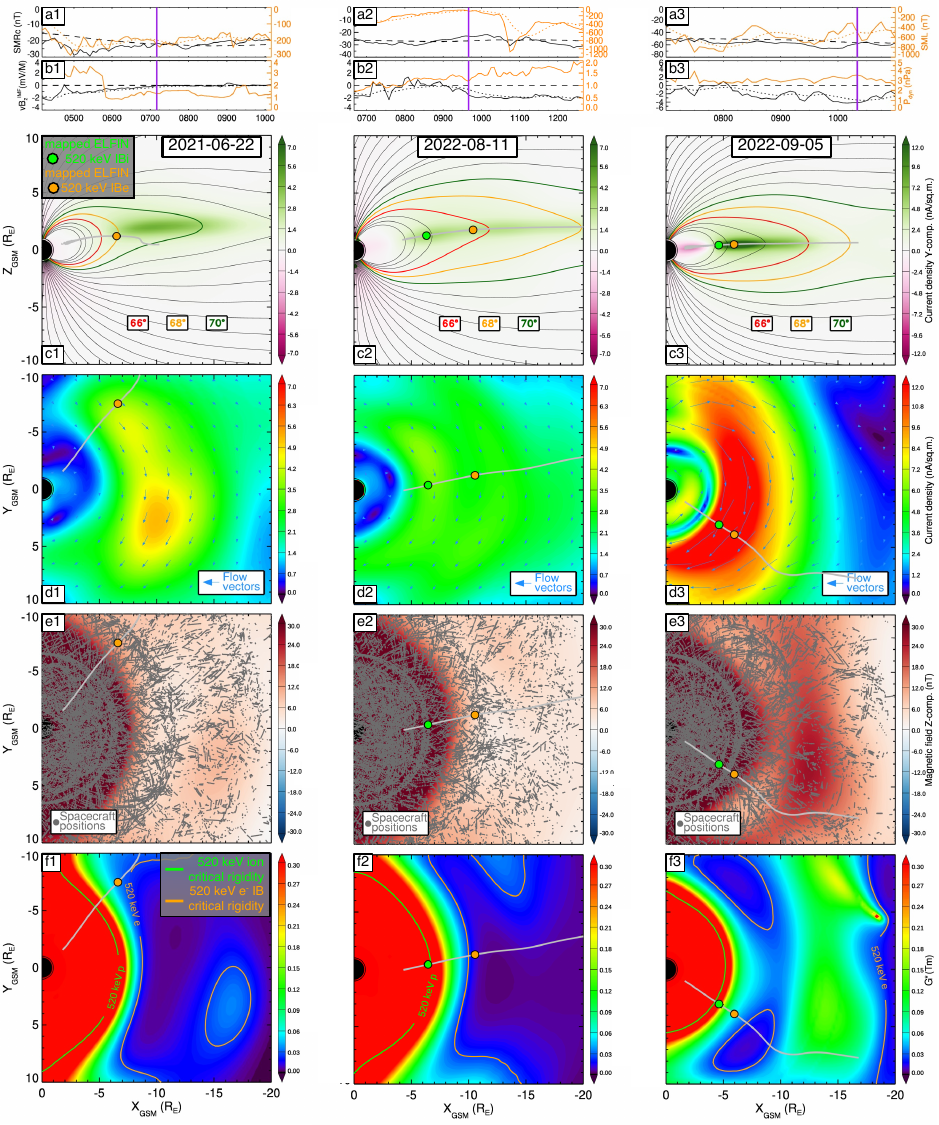}
  \caption{(Caption next page.)}
\label{figS9}
\end{figure}
\addtocounter{figure}{-1}
\begin{figure} [t!]
 \caption{SST19 reconstructions of IBs compared to mapped ELFIN observed locations for three more events. (a) The SuperMAG pressure-corrected storm index $\textit{SMRc}$ (black line) and substorm index $\textit{SML}$ (orange line). Their smoothed values, used in the $K_\mathrm{NN}$ procedure, are indicated by the dashed lines. (b) The solar wind electric field parameter $vB_z^\mathrm{IMF}$ (black line) and dynamic pressure $P_\mathrm{dyn}$ (orange line). The smoothed value of $vB_s^\mathrm{IMF}$, used in the $K_\mathrm{NN}$ procedure, is indicated by the dashed line. The reconstructed time is shown by the vertical purple line. (c--f) The panels are the same as Figures~S2 and S3, but for different events.}
\end{figure}

\end{document}